\documentclass[11pt]{article} 
\def\ds{\displaystyle}

\begin{document}
\pagestyle{headings}
\renewcommand{\thefootnote}{\alph{footnote}}

\title{Algebras without Involution and Quantum Field Theories}
\author{Glenn Eric Johnson\\Oak Hill, VA.\footnote{Author is reached at: glenn.e.johnson@gmail.com.}}
\maketitle

{\bf Abstract:} Explicit realizations of quantum field theory (QFT) are admitted by a revision to the Wightman axioms for the vacuum expectation values (VEV) of fields. The technical development of QFT is expanded beyond positive functionals on $*$-algebras while the physically motivated properties: Poincar\'{e} covariance; positive energy; microcausality; and a Hilbert space realization of states, are preserved.

\section{Introduction}


A precise description of quantum field theory (QFT) that both realizes the first principles of quantum mechanics and includes the successful phenomenology has been elusive. At issue is to either verify that proposed descriptions of QFT are sufficient by demonstrating realizations of physical interest, or to find revised descriptions that admit realizations. This note demonstrates that consideration of algebras other than $*$-algebras results in realizations of QFT that exhibit interaction. The approach is to examine explicit constructions of positive functionals that satisfy the Wightman axioms [\ref{wight},\ref{pct},\ref{bogo},\ref{borchers}] on a subalgebra of a $*$-algebra of function sequences. This subalgebra consists of functions with Fourier transforms that lack support on negative energies, and consequently the subalgebra lacks the involution of the enveloping $*$-algebra. Quantum mechanics and relativity are realized without a constraint that fields be Hilbert space operators.  A more general correspondence of field with state, one in analogy with cases from ordinary quantum mechanics, substitutes for self-adjoint Hilbert space field operators in the case of interacting fields.


The QFT dispute is illustrated by the lack of explicit four-point vacuum expectation values (VEV) for a scalar field.\footnote{There are demonstrations of realizations of QFT, notably for fields lacking interaction, fields not satisfying microcausality [\ref{locality},\ref{lechner}], and one dimensional space.} Significant efforts have not produced a realization for this generalized function required by non-trivial dynamics. Appropriate generalized functions have been characterized [\ref{gel1}], but the lack persists. In contrast, the elementary generalized function\begin{equation}\label{exampl}{^C \tilde{W}}_4((p)_4)=(2\pi)^dc_4\; \delta(p_1\!+\!p_2\!+\!p_3\!+\!p_4)\; \delta(p_1^2\!-\!m^2)\; \delta(p_2^2\!-\!m^2)\; \delta(p_3^2\!-\!m^2)\; \delta(p_4^2\!-\!m^2)\end{equation}realizes the connected component of a four-point VEV in a revised development. In three or more dimensions,\begin{equation}\label{expand4}\renewcommand{\arraystretch}{1} \begin{array}{rl} W_4((x)_4) &={^T W}_4((x)_4)+W_2(x_1,x_2)W_2(x_3,x_4)\\
 & \qquad \qquad +W_2(x_1,x_3)W_2(x_2,x_4) +W_2(x_1,x_4)W_2(x_2,x_3)\end{array}\end{equation}defines a generalized function (see Appendix A) that is evidently local and Poincar\'{e} invariant. Two-point functions for a neutral scalar field are described by K\"{a}ll\'{e}n-Lehmann forms [\ref{steinmann}] and two-point functions for any spin have been constructed [\ref{bogo},\ref{gel5}]. The singularity on mass shells results in scattering amplitudes, with a non-forward contribution\begin{equation} \langle (p_1,p_2)^{\mathit{out}} | (p_3,p_4)^{\mathit{in}} \rangle = (2\pi)^d c_4\, \delta(\omega_1 \!+\!\omega_2\!-\! 
\omega_3\!-\!\omega_4) \; \frac{\delta ({\bf p}_1 +{\bf p}_2 - {\bf p}_3 -{\bf p}_4)}{2^4 \omega_1\omega_2\omega_3\omega_4} \label{scatamp4}
\end{equation}that coincides up to a phase with the weak coupling result from the Feynman rules for a :$\Phi^4$: interaction [\ref{bjdrell2}] in the plane wave state normalization of this note.

Microcausality is an obstacle to constructions of non-trivial QFT [\ref{locality},\ref{lechner}]. The composition of local two-point functions and symmetric functionals (\ref{expand4}) satisfies locality but a symmetric functional (\ref{exampl}) does not have appropriate energy support. However, with symmetry broken by test functions of the form $f^* g$ with the Fourier transforms of $f$ and $g$ lacking support on negative energies,\footnote{The constraint on the test functions to support only positive energies eliminates compact spacetime support, familiar in the lack of ``localizable'' elementary states [\ref{wigner}]. The selected test functions are in a form $i\dot{f}+\sqrt{m^2-\Delta}f$ and both terms can not vanish in a domain unless $f=0$ [\ref{segal}]. Locality is verified in an enveloping set of test functions that include test functions of compact spacetime support and the states in the Hilbert space are labeled by the selected subset. The essential points of the VEV, and their sums and differences, are not modified in limiting the test functions used to construct the Hilbert space. The VEV are required to be generalized functions for the enveloping set of test functions.} and consequently the Fourier transform of $f^*$ lacking support on positive energies, closure of the cone $\bar{V}^+$ under addition of elements results in satisfaction of the spectral support condition. The revision to the original Wightman axioms is to limit the test functions used to construct states in the Hilbert space to those lacking support on the negative energy mass shells. The spectral support condition-violating symmetry of the VEV is broken by selection of Hilbert space states with positive energies. Limiting the test functions to those supported on positive energies is sufficient to demonstrate satisfaction of the spectral support condition. $p_4$ is limited to the closed forward cone $\bar{V}^+$ given that $p_4^2=m^2$ and the test functions support only positive energies. Then $p_3+p_4$ and $p_2+p_3+p_4$ are in $\bar{V}^+$ either as a consequence of closure of $\bar{V}^+$ with summation or $p_1+p_2+p_3+p_4=0$ implies that these vectors are the negatives of vectors in the closed backward cone $\bar{V}^-$.

For (\ref{exampl}), a Wightman semi-norm results from\[\renewcommand{\arraystretch}{1.5} \begin{array}{rl} {^T W}_4(f_2^* f_2)&=c_4{\ds \int} du\; \left|{\ds \int} d\sigma_m(p_1)d\sigma_m(p_2)\; e^{iu(p_1+p_2)}\tilde{f}_2(p_1,p_2) \right|^2\\ &\geq 0\end{array}\]with $d\sigma_m(p)=\delta(p^2-m^2)\,dp$, $c_4>0$ and $W_4(f_1^* g_3)=0$ for the energy support limited functions. Then, for the energy support limited test functions and up to VEV order four, (\ref{expand4}) satisfies the physical Wightman conditions. A revision to the Wightman axioms is necessary to consider these elementary VEV (\ref{exampl}) that are excluded by a full application of the axioms. 


The constructions exhibit properties: Poincar\'{e} covariance; positive energy; microcausality; particle creation and annihilation; and a Fock-like Hilbert space of states, that make them interesting as relativistic quantum mechanics. It is also interesting that prevalent assumptions in QFT do not apply: fields of physical interest are not Hermitian operators in the Hilbert space of states, and there are states that do not converge to states of the elementary particle's free field at large times. Although the constructed fields share many properties conjectured for Hilbert space field operators, the constructions do not satisfy the full set of conventional assumptions for QFT and consequently many well-known theorems no longer apply since their assumptions are not included within the revised axioms.\footnote{Since the non-trivial constructions lack Hermitian field operators, the Jost-Schroer and related theorems [\ref{bogo},\ref{mund}] do not imply that the two-point function must be distinct from that of a free field; and since the support of the constructed states is not localized in bounded regions (the support can be dominantly within bounded regions but does not vanish in any region), paradoxes for type I realizations and a Hilbert space of type III [\ref{type3}] no longer necessarily follow.} In particular, the presumption that the field is a symmetric\footnote{Use of the terms Hermitian, symmetric, hypermaximal-symmetric, and self-adjoint varies. Here the terms are that an operator $A$ with domain ${\cal D}_A$ in a Hilbert space with scalar product $\langle u,v \rangle$ is {\em Hermitian} if $\langle u,Av\rangle= \langle Au,v \rangle$ for every $u,v \in {\cal D}_A$, a Hermitian operator is {\em symmetric} if ${\cal D}_A$ is dense, and a symmetric operator is {\em self-adjoint} if ${\cal D}_A={\cal D}_{A^*}$ and $Au=A^*u$ for every $u\in {\cal D}_A$. Then self-adjoint implies symmetric implies Hermitian.} operator and equivalent assertions are abandoned. The free field and related constructs result in symmetric field operators but these are special cases with VEV based on the Pauli-Jordan two-point function. Here, conjecture on whether a field is a Hermitian Hilbert space operator is removed from the foundations of QFT.

Positive functionals on tensor algebras of function sequences provide realizations of QFT. Two approaches to Algebraic QFT [\ref{borchers},\ref{yngvason}] are distinguished here:\begin{itemize} \item[1)] Describe algebras of local, self-adjoint operators with Hilbert space representations that exhibit physical properties. \item[2)]  Describe algebras of the function sequence labels of elements of Hilbert spaces that exhibit physical properties.\end{itemize}Here, ``physical properties'' refers to: Poincar\'{e} invariant amplitudes; positive energy spectra for the generators of translations; and local VEV. Motivated by the apparent difficulty [\ref{summers}] of achieving interaction in four spacetime dimensions for 1), the approach studied here is 2). 2) is expressed with fewer assumptions and has realizations for situations of physical interest.

Hermitian field operators are no longer a consequence of a positive functional satisfying the Wightman axioms when the algebra of function sequences lacks the $*$-involution of a Borchers-Uhlmann algebra. An algebra of function sequences without support on negative energies lacks the $*$-involution. Without the $*$-involution, the null space for the Wightman semi-norm is not necessarily a left ideal and representation of the algebra in the Hilbert space no longer necessarily follows. Representation of the algebra in the Hilbert space would define the field as a Hermitian operator but in the revised context, generally the mappings in the $*$-algebra of function sequences anticipated to result in field operators do not associate with positive energy spectra, and when restricted to positive energy, adjoints of the fields map elements out of the Hilbert space. The definition for a field follows the algebraic QFT development except for representation by Hermitian Hilbert space operators. A quantum mechanical model results from Hilbert space representation of the linear vector space properties of the algebra of function sequences.

The assertion that fields are Hermitian Hilbert space operators is a well established convention and an extrapolation from both ordinary quantum mechanics and free fields but self-adjointness of an associated Hilbert space operator is not necessary for a field to be manifest in the classical limits of states. In interpretations of quantum mechanics that eliminate ``collapse of the wave packet''\footnote{Collapse of wave packets onto the states in the ranges of the orthogonal projections in an expansion of a self-adjoint operator associates ``observables'' with self-adjoint operators.} such as the Everett interpretation [\ref{ewg}], entanglement [\ref{entangle}] of states constitutes measurement, that is, a measurement is any interaction that entangles system states with observer states. Questions of interpretation in quantum mechanics are without empirical consequence when both the mathematical formulation and interpretation of values are agreed upon. However, assertion that fields are Hermitian Hilbert space operators, that is, that only properties associated with orthogonal projections are observable, excludes these QFT constructions for the cases of physical interest. Fields are observable quantities, but it is a question of interpretation to distinguish observable, meaning susceptible to measurement, from the definition of an ``observable'' as a self-adjoint operator. In rigged Hilbert spaces (Gelfand triplets), the resolutions of unity determine the self-adjoint operators. The kernel theorem [\ref{gel4}] provides that self-adjoint operators are linear combinations with real coefficients of the projection operator elements of a resolution of unity. While decomposition in orthogonal projections occurs in important cases of physical interest, failure of self-adjointness for a corresponding operator does not exclude a property from the descriptions of states nor prevent entanglement of observer with system. Interpretation of a measurement as due to a property of the interacting system rests on that the description of the state includes the property, but not that the property correspond with orthogonal projections. With Lorentz invariance, position is a property of particle states that lacks a corresponding orthogonal projection onto eigenstates of position [\ref{wigner}].\footnote{The generally complex $\langle f|x f\rangle= -i\int dp \;(\delta(E-\omega)/2\omega)\; \overline{\tilde{f}(p)}\, (\partial \tilde{f}(p)/\partial p)$ are not components of position except in the non-relativistic, classical limit $m\rightarrow \infty$.}

Lack of a Hermitian operator corresponding to an observable quantity is familiar in other cases of physical interest. Ordinary quantum mechanics provides the examples $x^n p$, quantities that are undistinquished in a classical limit but do not correspond to Hermitian operators. The operator associated with $x^n p$ exhibits eigenvectors with imaginary eigenvalues in ${\cal L}_2$.\footnote{The operator $X^{n/2}PX^{n/2}$ has eigenvectors $e_\lambda (x)= x^{-n/2} \exp(-\lambda/((n-1)x^{n-1}))$ with eigenvalues $-i\hbar \lambda$ for $x,\lambda >0$, $n\geq2$. From the Riesz-Fischer theorem, there are resolutions of unity in the generalized eigenvectors of $X$ or $P$ for ${\cal L}_2$ that distinguish $X$ and $P$ as ``observables'' in ordinary quantum mechanics. But, due to Lorentz invariance, $X$ is not similarly distinguished in relativistic physics.} The formally Hermitian $\frac{1}{2}(X^nP+PX^n)=X^{n/2}PX^{n/2}$ is not a Hermitian ${\cal L}_2$ operator. Nevertheless, for particular states $|s(t)\rangle$ such as minimum uncertainty packet states with small spatial variances, the trajectory of $x^np$ given by Newtonian mechanics approximates the $\langle s(t)| X^{n/2}PX^{n/2} s(t)\rangle$ given by quantum dynamics [\ref{schrodinger}]. This establishes that there are many states that have real $\langle s(t)| X^{n/2}PX^{n/2} s(t)\rangle$ that agree with classical limits $x^np$ while $X^{n/2}PX^{n/2}$ is not self-adjoint. Without additional conditions, an assertion that observable properties of states correspond with self-adjoint operators is unjustified.

The revised axioms depart from the historical emphasis of QFT on canonical quantization of classical field equations. Study of dynamical descriptions consistent with realizations of the Wightman or Araki-Haag-Kastler axioms is potentially more general than particular dynamical models, and indeed, in the constructed Hilbert spaces, demonstrably non-trivial dynamics result from ``trivial'' Hamiltonians. There is inherent interest to not limiting quantum dynamics to that derived from classical limits (albeit, constrained to agree). It is not evident that ``quantization'' of dynamical laws based on the classical concepts of state evolution as trajectories of labeled particles suffices for relativistic quantum dynamics. This study also does not assert that there are local algebras of self-adjoint operators but rather focuses on realization of QFT as quantum mechanics, that is, realizations of states as elements in Hilbert spaces with scalar products that exhibit the physical properties. Considering the lack of physically relevant realizations of the original axioms and with (\ref{exampl}) available from [\ref{agw},\ref{iqf}], the conjectures equivalent to requiring Hermitian field operators are abandoned. While this revision deviates significantly from common practices of QFT, satisfaction of the physically motivated Wightman axioms and scattering amplitudes that match contributions from Feynman series display promise. It is demonstrated that this revised development admits realizations for any finite spin and justifies a revisit to dynamical laws and their classical limits.

\section{Definitions and notation}

The constructions are in $d \geq 3$ spacetime dimensions with masses $m_\kappa>0$. The spacetime coordinates $x$ and energy-momentum vectors $p$ are $x:=t,{\bf x}$, $p:=E,{\bf p}$ and more generally $q:=q_{(0)},{\bf q}$ with $x^2:=x^T g x=t^2-{\bf x}^2$ for the Minkowski signature $g$. $x,p,q\in {\bf R}^d$, ${\bf x},{\bf p},{\bf q}\in {\bf R}^{d-1}$, ${\bf x}^2:=x_{(1)}^2\!+\!\ldots x_{(d-1)}^2$ is the square of the Euclidean length in ${\bf R}^{d-1}$, and $E_j^2=\omega_j^2:=m_{\kappa_j}^2+{\bf p}_j^2$ describe mass shells. $p \in \bar{V}^+$, the closed forward cone, if $p^2 \geq 0$ and $E\geq 0$ (backward cone has $E\leq 0$). Multiple variables are denoted by $(x)_n:=x_1,x_2\ldots x_n$ and $(x)_{k,n}:=x_k,\ldots x_n$ for either ascending or descending sequences of indices. Repetition of variables includes recursion, for example,\begin{equation}\label{recursion}(\sum_{\nu}(\int d\zeta)_2)_3:= \sum_{\nu_1} \int d\zeta_1 \int d\zeta_2 \sum_{\nu_2} \int d\zeta_3 \int d\zeta_4 \sum_{\nu_3} \int d\zeta_5 \int d\zeta_6.\end{equation}Dirac delta generalized functions supported on mass shells are denoted\[\renewcommand{\arraystretch}{1.25} \begin{array}{rl} \delta_j^{\pm}&:= \delta(\pm E_j - \omega_j)/(2\omega_j)\\
\hat{\delta}_j &:= \delta(p_j^2-m_{\kappa_j}^2) = \delta_j^+ + \delta_j^-.\end{array}\]Sign conventions for Fourier transforms are set by the test functions,\[ \tilde{f}_n((p)_n) := \frac{1}{\sqrt{(2 \pi)^{nd}}}\,\int_{{\bf R}^{nd}} (dx)_n \; \prod_{k=1}^n e^{-ip_k x_k} f_n((x)_n),\]together with the definition of the Fourier transform of generalized functions $\tilde{T}(\tilde{f}) = T(f)$ and $p_jx_k:=p_j^T gx_k=E_jt_k-{\bf p}_j\cdot {\bf x}_k$. $\overline{\alpha}$ denotes the complex conjugate of $\alpha$. Summation notation is used for generalized functions, $\int dx\; T(x) f(x):=T(f)$.

\subsection{Algebraic Quantum Field Theory and test functions}

A Borchers-Uhlmann algebra consists of terminating sequences [\ref{yngvason}]\[ \underline{f} := (\, f_0,\ldots, f_n((x)_n)_{\kappa_1\ldots \kappa_n} ,\ldots).\]Each $f_n((x)_n)_{\kappa_1\ldots \kappa_n}$ is one of a sequence of $(N_c)^n$ test functions with $(x)_n\in {\bf R}^{nd}$ and each $\kappa_k\in \{1,\ldots N_c\}$. $f_0$ is a complex number. The $\underline{f}$ have a topology inherited from the component functions. In [\ref{borchers}], the algebra was denoted $\Sigma$ with component functions that are Schwartz, tempered test functions, $S({\bf R}^{nd})$. For the constructions, the algebra is designated ${\cal A}$ and has component energy-momentum functions $\tilde{f}_n((p)_n)_{\kappa_1 \ldots \kappa_n}$ that are the multiple argument versions of the span of products of a tempered test function $\tilde{f}({\bf p}) \in S({\bf R}^{d-1})$ and a multiplier $\tilde{g}(p)$ in $S({\bf R}^d)$ [\ref{gel2}].\[\tilde{f}_1(p)=\tilde{g}(p) \tilde{f}({\bf p})\in {\cal A}.\]The component functions of ${\cal A}$ are test functions of $({\bf p})_n$ when evaluated on mass shells, $\tilde{f}_n((\pm\omega,{\bf p})_n)_{\kappa_1 \ldots \kappa_n} \in S({\bf R}^{n(d-1)})$. $\Sigma \subset {\cal A}$. The constructed generalized functions have the form\[\int dE\, d{\bf p}\; \delta(E\pm \omega) T({\bf p}) \tilde{f}_1(p)=\int d{\bf p}\; T({\bf p}) \tilde{f}_1(\pm \omega, {\bf p})\]for each of the multiple arguments and components, and $T({\bf p})\in S'({\bf R}^{d-1})$, a generalized function. A Fourier transform of $\tilde{f}_n((p)_n)_{\kappa_1 \ldots \kappa_n}$ is the generalized function $f_n((x)_n)_{\kappa_1 \ldots \kappa_n}$ in $S'({\bf R}^{nd})$ given by $(f_n,\psi_n)=(\tilde{f}_n,\tilde{\psi}_n)$ for $\psi_n((x)_n)\in S({\bf R}^{nd})$. These spacetime functions are equivalent to test functions when $\tilde{f}_1(p) \in \Sigma$, and proportional to derivatives of $\delta(t)$ times spatial test functions when $\tilde{f}_1(p)$ is $E^k\tilde{f}({\bf p})$.

${\cal B}$ is a subset of ${\cal A}$ that vanishes on negative energy mass shells. For every $f_n \in {\cal A}$, let\[ \tilde{\varphi}[ f_n]((p)_n)_{\kappa_1 \ldots \kappa_n} := \prod_{k=1}^n (\omega_k + E_k) \tilde{f}_n((p)_n)_{\kappa_1 \ldots \kappa_n}.\]and $\varphi[1] := 1$. $\varphi[{\cal A}]\subseteq {\cal B}$. $\omega_k$ and $E_k$ are multipliers so $\underline{W}(\varphi[\underline{f}])$ is bounded when $\underline{W}(\underline{f})$ is bounded. LSZ states, used to define scattering amplitudes in QFT, appear naturally in this development. States labeled by functions based upon\[\tilde{\ell}(t;p_k):=(\omega_k + E_k) e^{i\omega_k t} \tilde{f}({\bf p}_k),\]are LSZ states and are elements of ${\cal B}$. $t$ is a parameter of the LSZ function.

Given a sequence of generalized functions $\underline{W}$, interpretation of $\underline{W}$ as VEV of fields derives from a sesquilinear function on ${\cal A}\times {\cal A}$.\begin{equation}\label{sesqui} \renewcommand{\arraystretch}{1.5} \begin{array}{rl} \langle \underline{f}| \underline{g} \rangle &:= {\ds \sum_{n,m} \;\sum_{\kappa_1=1}^{N_c}\ldots \sum_{\kappa_{n+m}=1}^{N_c} \int} (dp)_{n+m} \; \tilde{W}_{n+m}((p)_{n\!+\!m})_{\kappa_1\ldots \kappa_{n\!+\!m}}\times \\
 & \qquad \qquad \qquad \qquad \qquad \tilde{f}_n^*((p)_n))_{\kappa_1 \ldots \kappa_n} \; \tilde{g}_m((p)_{n+1,n+m})_{\kappa_{n\!+\!1} \ldots \kappa_{n\!+\!m}}.\end{array}\end{equation}To shorten the notation, let $(\xi)_{I_n}:=(p,\kappa)_{I_n}$, $(-\xi)_{I_n}:=(-p,\kappa)_{I_n}$ and\[\int (d\xi)_{I_n}:=\sum_{\kappa_{i_1}=1}^{N_c}\ldots \sum_{\kappa_{i_n}=1}^{N_c} \int dp_{i_1}\ldots dp_{i_n}\]for any set of indices $I_n=\{i_1,i_2,\ldots i_n\}$. Then $\tilde{f}_n((\xi)_n):=\tilde{f}_n((p)_n)_{\kappa_1 \ldots \kappa_n}$ and relabeling summations results in\begin{equation} \renewcommand{\arraystretch}{1.25} \begin{array}{rl} \langle \underline{f}| \underline{g} \rangle &= {\ds \sum_n \;\int} (d\xi)_n\; \tilde{W}_n((\xi)_n) {\ds \sum_{\ell=0}^n} \tilde{f}_\ell^*((\xi)_{\ell})\; \tilde{g}_{n-\ell}((\xi)_{\ell+1,n})\\
&=\underline{W}(\underline{f}^* \,{\bf x}\, \underline{g})\end{array} \label{th1} \end{equation}with the product\begin{equation} \underline{f} \,{\bf x}\, \underline{g} := (f_0g_0, \ldots, \sum_{\ell=0}^n f_{\ell}((x)_{\ell})_{\kappa_1\ldots \kappa_{\ell}} \,g_{n-\ell}((x)_{\ell+1,n})_{\kappa_{\ell+1}\ldots \kappa_n}, \dots),\label{prod}\end{equation}and the $*$-map\begin{equation}\renewcommand{\arraystretch}{2} \begin{array}{rl} \tilde{f}_n^*((\xi)_n) &:= ((D^T \cdot)_n \overline{\tilde{f}_n} ((-\xi)_{n,1}))\\
 &= {\ds \sum_{\ell_1} \ldots \sum_{\ell_n}}\; D_{\ell_1 \kappa_1}\ldots D_{\ell_n \kappa_n} \overline{\tilde{f}_n} (-p_n, -p_{n-1}, \ldots ,-p_1)_{\ell_n \ldots \ell_1}.\end{array} \label{dualf}\end{equation}

Considering sums and products of the fields $\Phi(x)_{\kappa}$ as elements of a ring ${\cal R}$ over the complex numbers, the Wightman functions are the VEV of products of fields.\begin{equation}\label{sesquis2} \renewcommand{\arraystretch}{1.25} \begin{array}{rl} W_n((x)_n)_{\kappa_1 \ldots \kappa_n} &:= \langle \Omega| \Phi(x_1)_{\kappa_1}\ldots \Phi(x_n)_{\kappa_n} \Omega\rangle\\
 &\;= \langle \Phi(x_k)_{\kappa_k} \ldots \Phi(x_1)_{\kappa_1}\Omega| \Phi(x_{k+1})_{\kappa_{k+1}}\ldots \Phi(x_n)_{\kappa_n} \Omega\rangle \end{array}\end{equation}independently of $k$. $\underline{f}\in {\cal A}$ is associated formally with an element of ${\cal R}$.\begin{equation}\label{elements} u(\underline{f}) := {\ds \sum_n \sum_{\kappa_1=1}^{N_c}\ldots \sum_{\kappa_n=1}^{N_c} \int} (dx)_n\;f_n(x_1, \ldots x_n)_{\kappa_1\ldots \kappa_n} \Phi(x_1)_{\kappa_1}\ldots \Phi(x_n)_{\kappa_n}.\end{equation}For $\underline{f}=(0,\ldots 0,f(x),0,\ldots)$, $f(x)$ corresponding to $\Phi(x)_\kappa$ in (\ref{elements}), (\ref{prod}) identifies the field as multiplication of function sequences, $\Phi(f)_\kappa\, \underline{g} = \underline{f} \,{\bf x}\, \underline{g}$.\[\renewcommand{\arraystretch}{1.5} \begin{array}{rl} u(\underline{f}\, {\bf x}\, \underline{g}) &= {\ds \sum_n \sum_{\kappa_1=1}^{N_c}\ldots \sum_{\kappa_n=1}^{N_c} \int} dx' (dx)_n\;f(x')\, g_n(x_1, \ldots x_n)_{\kappa_1\ldots \kappa_n}\, \Phi(x')_\kappa \Phi(x_1)_{\kappa_1}\ldots \Phi(x_n)_{\kappa_n}\\
&= {\ds \sum_n \sum_{\kappa_1=1}^{N_c}\ldots \sum_{\kappa_n=1}^{N_c} \int} (dx)_n\; g_n(x_1, \ldots x_n)_{\kappa_1\ldots \kappa_n} \Phi(f)_\kappa \Phi(x_1)_{\kappa_1}\ldots \Phi(x_n)_{\kappa_n}\\
&= \Phi(f)_\kappa \;u(\underline{g}). \end{array} \]Should the field be an operator in the constructed Hilbert space, then fields are designated Hermitian if for $\underline{f}=\underline{f}^*$\[\Phi^*(f)=\Phi(f)\]for elements in a common domain and with $\Phi(f)=\sum_\kappa \Phi(f_\kappa)_\kappa$. The adjoint is given by\begin{equation}\label{hermitian}\renewcommand{\arraystretch}{1.5} \begin{array}{rl}
\langle \underline{h} | \Phi(f)\underline{g} \rangle&= \underline{W}(\underline{h}^* \,{\bf x}\, (\underline{f} \,{\bf x}\, \underline{g}))\\
 &= \underline{W}((\underline{f}^* \,{\bf x}\, \underline{h})^* \,{\bf x}\, \underline{g}). \end{array}\end{equation}Then $\Phi^*(f)= \Phi(f^*)$ is the adjoint operator. The field components are formally Hermitian, $\Phi^*(x)_\kappa= (D\Phi(x))_\kappa$, from (\ref{th1}), (\ref{dualf}) and (\ref{sesquis2}) but when there are only trivial $f=f^*$ then the formally Hermitian fields are not Hermitian. Should there be an operator $\Phi(f)$, it is not Hermitian for the non-trivial constructions since $f= f^* \in {\cal B}$ implies that $\| \underline{f} \| =0$. Such $\underline{f}$ are in a two-sided ideal. 

The nonsingular linear transformation $D$, Dirac conjugation, is selected to satisfy\begin{equation}\label{cond-D} \overline{D}D=1.\end{equation}Then the $*$-map (\ref{dualf}) satisfies\begin{equation} \renewcommand{\arraystretch}{1.25} \begin{array}{rl} \underline{f}^{**}&=\underline{f}\\
(\lambda \underline{f})^* &=\overline{\lambda} \underline{f}^* \\
(\underline{g}+\underline{f})^* &=\underline{g}^* + \underline{f}^*\\
(\underline{g} \,{\bf x}\, \underline{f})^* &=\underline{f}^* \,{\bf x}\, \underline{g}^*. \end{array} \end{equation}This $*$-map is an involution [\ref{emch}] when it is an automorphism. This $*$-map (\ref{dualf}) is an automorphism of ${\cal A}$ but not of ${\cal B}$.

\subsection{Conditions on $M(p)$} The Wightman-functional for a free field $\underline{W_o}$ is described by a two-point functional. The two-point functional used in the constructions is set equal to this free field two-point function.\footnote{By the Jost-Schroer and related theorems, a field can not be a local Hermitian Hilbert space operator if the interaction is non-trivial and the two-point function has this form [\ref{bogo},\ref{mund}]. For the constructions, the field is not a Hermitian operator for models of physical interest.} The Fourier transform of the two-point function is\begin{equation}\renewcommand{\arraystretch}{1.25} \begin{array}{rl} \tilde{\Delta}(p_1,p_2)_{\kappa_1 \kappa_2} &:= \delta(p_1+p_2)\; \delta_2^+ \, M_{\kappa_1 \kappa_2}(p_2)\\
&= \delta({\bf p}_1+{\bf p}_2)\; \delta_1^- \delta_2^+ \, 2\sqrt{\omega_1\omega_2}\, M_{\kappa_1 \kappa_2}(p_2).\end{array}\label{twopoint}\end{equation}The $M(p)_{\kappa_1 \kappa_2}$ are multinomials in energy-momentum components. For $\underline{f},\underline{g} \in {\cal B}$,\begin{equation}\label{twoptsupp} \Delta(f_2)_{\kappa_1 \kappa_2}=\Delta(f_2^*)_{\kappa_1 \kappa_2} =\Delta(f_1 g_1^*)_{\kappa_1 \kappa_2}=0,\end{equation}consequences of the lack of support at negative energies. The condition\begin{equation} \label{matcond} DM(p) = C^*(p) C(p) \end{equation}results in the semi-norm $\|\underline{f}\|_o:=\sqrt{\underline{W_o}(\underline{f}^* \,{\bf x}\, \underline{f})}$. $D$ is from (\ref{dualf}). Since $DM(p)$ is nonnegative, it is Hermitian and\[\renewcommand{\arraystretch}{1.25} \begin{array}{rl}
DM(p) &= (DM(p))^*\\
&= M(p)^* \overline{D^T}\\
&= M(p)^* (D^T)^{-1} \end{array} \]using (\ref{cond-D}). Then, for these constructions\begin{equation} M(p)^*=DM(p) D^T.\label{Mstar}\end{equation}

$M(p)$ that have direct sum decompositions [\ref{horn}] into components $M_k(p)$,\[M= \left( \renewcommand{\arraystretch}{1} \begin{array}{cc} M_1 & 0\\ 0 & M_2\end{array} \right),\]and that satisfy locality conditions\begin{equation} \label{matcond0} M_k(-p)^T = \pm M_k(p)\end{equation}are used in the constructions. In (\ref{matcond0}), the $N_b \times N_b$ boson component $M_1$ uses `$+$' and the fermion component $M_2$ uses `$-$'. The convention here is that $\kappa \in \{1,2,\ldots N_b\}$ are boson field components and $\kappa \in \{N_b+1,\ldots N_c\}$ are fermion field components.

Amplitudes (\ref{th1}) are invariant under Poincar\'{e} transformations.\begin{equation} \label{innerp} \underline{W}(\underline{f}^*\,{\bf x}\, \underline{g}) =  \underline{W}(((a,A)\underline{f})^*\,{\bf x}\, (a,A)\underline{g})\end{equation}with\[(a,A)\underline{f}:=(f_0,\ldots ((S(A)^T\cdot)_n f_n((\Lambda^{-1}(x\!-\!a))_n))_{\kappa_1 \ldots \kappa_n},\ldots).\]In the constructions, two conditions\begin{equation} \label{matcond2} \renewcommand{\arraystretch}{1.25} \begin{array}{rl}
S(A)M(p)S(A)^T &= M(\Lambda^{-1} p)\\
\overline{S}(A) D&= DS(A) \end{array} \end{equation}result in Poincar\'{e} covariance,\begin{equation}\label{p-cov}((S(A)\cdot)_{n+m} W_{n+m}((x)_{n+m}))_{\kappa_1 \ldots \kappa_{n+m}}=W_{n+m}((\Lambda^{-1}(x-a))_{n+m})_{\kappa_1 \ldots \kappa_{n+m}}.\end{equation}Example realizations of $M(p), D, S(A)$ are provided in Appendix B.

\subsection{Symmetrization} Summation over signed permutations of arguments is denoted by\begin{equation} {\bf S}[T_n((x)_n)_{\kappa_1 \ldots \kappa_n}] := \sum_{\pi} s_{\kappa_{\pi_1} \ldots \kappa_{\pi_n}} \,T_n(x_{\pi_1},x_{\pi_2},\ldots x_{\pi_n})_{\kappa_{\pi_1} \ldots \kappa_{\pi_n}}. \label{permutation}\end{equation}The summation includes all $n!$ permutations of $\{1,2,\ldots n\}$. The signs $s_{\kappa_{\pi_1} \ldots \kappa_{\pi_n}}$ are determined by transpositions,\begin{equation}s_{\ldots \kappa_j \kappa_{j+1}\ldots }= \sigma_{\kappa_j \kappa_{j+1}} s_{\ldots \kappa_{j+1} \kappa_j\ldots }\label{signs-perm}\end{equation}with $\sigma_{\kappa_j \kappa_{j+1}}=- 1$ if $\kappa_j, \kappa_{j+1} > N_b$, and $\sigma_{\kappa_j \kappa_{j+1}}= 1$ otherwise. These signs are set to agree with the commutation relations of the free field that apply when $x_i-x_j$ is space-like.\[\Delta(x_i,x_j)_{\kappa_i \kappa_j}= \sigma_{\kappa_i \kappa_j} \Delta(x_j,x_i)_{\kappa_j \kappa_i}\]when $(x_i-x_j)^2 <0$. The signs $s_{\kappa_{\pi_1} \ldots \kappa_{\pi_n}}$ are determined by transpositions relative to one sign. The argument of ${\bf S}[\cdot]$ indicates a term with positive sign.

Summations over arbitrary subsets of arguments are designated\begin{equation} {\bf S}_{\chi_1^k}[T_n((x)_n)_{\kappa_1 \ldots \kappa_n}] := \sum_{\pi} s_{\kappa_{\pi_1} \ldots \kappa_{\pi_k}} \,T_n(x_{\pi_1},x_{\pi_2},\ldots x_{\pi_k},x_{k+1},\ldots x_n)_{\kappa_{\pi_1} \ldots \kappa_{\pi_k}\kappa_{k+1} \ldots \kappa_n} \label{perm-sub}\end{equation}for the example of $\chi_1^k:=\{1,2,\ldots k\}$.

The $*$-map of a permuted sequence is the same as the permutation of the $*$-map.\begin{equation}\label{sdual}{\bf S}[\underline{f}]^*={\bf S}[\underline{f}^*].\end{equation}This is verified by noting that with the possible exception of sign the terms agree since both are sums over the permutations. If the sign of one term of ${\bf S}[\underline{f}]^*$ agrees with the corresponding term from ${\bf S}[\underline{f}^*]$ then the signs, set by transpositions, agree. From ${\bf S}[\underline{f}]^*$, the sign of $T_n((x)_n)_{\kappa_1 \ldots \kappa_n}$ is positive and the $*$-map results in $((D\cdot)_n \overline{T_n}((x)_{n,1})_{\kappa_n \ldots \kappa_1}$. By definition, the sign of $((D\cdot)_n \overline{T_n}((x)_{n,1})_{\kappa_n \ldots \kappa_1}$ from ${\bf S}[\underline{f}^*]$ is positive.

\section{Revising the Wightman axioms}

To allow the possibility that fields are not Hermitian operators, the original Wightman axioms are revised. Hermitian field operators are a consequence of the Wightman axioms for VEV [\ref{wight},\ref{pct},\ref{bogo},\ref{borchers}]. In the revision, the physically motivated axioms are maintained but the requirement that the Hilbert space construction uses an algebra of function sequences with the involution (\ref{dualf}) is eliminated. Inclusion of the involution is implicit in the application of the Wightman conditions for all tempered functions. This revision to the Wightman axioms includes both the constructions and established developments. In the revision, $\underline{W}$ is called a Wightman-functional and defines a quantum field theory if it has the following properties:

\begin{itemize} \item[A.I] $\underline{W}$ is a continuous linear functional for a Borchers-Uhlmann algebra ${\cal U}$.
\item[A.II] $\underline{W}$ is invariant under the proper orthochronous Poincar\'{e} group. $\underline{W}(\underline{f}) = \underline{W}((a,A) \underline{f})$ using the automorphism of ${\cal U}$ defined by\[(a,A)\underline{f}:=(f_0, \ldots ,((S(A)^T\cdot)_n\, f_n(\Lambda^{-1}(x_1-a), \ldots \Lambda^{-1}(x_n-a)))_{\kappa_1 \ldots \kappa_n},\ldots )\]with $S(A)$ an $N_c$-dimensional representation of the universal covering group of the proper orthochronous Lorentz group, $A\in \mbox{SL}(2)$, $\Lambda=\Lambda(A)$, and $a\in {\bf R}^d$.
\item[A.III] $\underline{W}$ is supported on positive energies. $\underline{W}({\cal M}_{sp})=0$ for ${\cal M}_{sp}$ the linear subspace of ${\cal U}$ with a base of functions $f_n((x)_n)_{\kappa_1\ldots \kappa_n}$ composed of products\[f_n((x)_n)_{\kappa_1\ldots \kappa_n}=u^*((x)_k)_{\kappa_1\ldots \kappa_k} v((x)_{k+1,n})_{\kappa_{k+1}\ldots \kappa_n}\]with $u, v \in {\cal V}\subseteq {\cal U}$ and with Fourier transforms $\tilde{f}_n((p)_n)_{\kappa_1\ldots \kappa_n}$ that vanish together with their derivatives when $p_n,p_{n-1}+p_n,\ldots, p_2+\ldots p_n \in \bar{V}^+$ and $p_1+p_2+\ldots p_n=0$.
\item[A.IV] $\underline{W}$ is local. $\underline{W}(I_c)=0$ with $I_c$ the linear subspace of ${\cal U}$ with a base of functions $f_n((x)_n)_{\kappa_1 \ldots \kappa_n}$ that decompose in the difference\[f_n =g((x)_n)_{\kappa_1\ldots \kappa_n}-P_{\pi;i,k}\, g((x)_n)_{\kappa_1\ldots \kappa_n}\]with the permutation of arguments defined by\[ P_{\pi;i,k}\, g((x)_n)_{\kappa_1\ldots \kappa_n} := \pm g(x_1\dots x_{i-1},x_{\pi_i} \ldots x_{\pi_k}, x_{k+1}\ldots x_n)_{\kappa_1\dots \kappa_{i-1},\kappa_{\pi_i} \ldots \kappa_{\pi_k}, \kappa_{k+1}\ldots \kappa_n}\]with $g=0$ if $x_j-x_{\ell}$ is time-like for $j\neq \ell$; $j,\ell \in \{i,i+1,\ldots k\}$. $\pi_i\dots \pi_k$ is any permutation of $i\ldots k$, and the sign is determined by (\ref{signs-perm}).
\item[A.V] $\underline{W}$ is a nonnegative functional. $\underline{W}(\underline{f}^* \, {\bf x}\, \underline{f})\geq 0$ for $\underline{f}\in {\cal V} \subseteq {\cal U}$.\end{itemize}(\ref{dualf}) defined the shorthand $((S \cdot)_n \,f_n ((x)_n))_{\kappa_1 \ldots \kappa_n}$.

A.1-A.V are a revision of the Wightman axioms that includes realizations of interest. A Borchers-Uhlmann algebra is involutive, ${\cal U}={\cal U}^*$, and the elements of ${\cal V}\subseteq {\cal U}$ determine the physically allowed states. If ${\cal V}={\cal U}=\Sigma$, the algebra of sequences derived from Schwartz's space of tempered functions, then these are multiple component field versions of Borchers' statement [\ref{borchers}] of the Wightman axioms. For the constructions, ${\cal V}={\cal B}\subset {\cal U}={\cal A}$ and A.III and A.V are revised from the original axioms to apply to elements of the subalgebra ${\cal B}$, functions with Fourier transforms supported only on positive energy. Continuity, Poincar\'{e} covariance, and locality still apply in ${\cal A}$. ${\cal B}$ is an algebra, closed under orthochronous Poincar\'{e} transformations but ${\cal B}$ is not involutive, ${\cal B}\neq {\cal B}^*$. ${\cal B}$ includes functions with Fourier transforms of compact energy-momentum support and $\underline{1}=(1,0,\ldots)\in {\cal B}$. Limiting the test functions is the modification to the Wightman axioms that enables implementation of the spectral support conditions for the elementary VEV (\ref{exampl}) and a Euclidean region semi-norm [\ref{iqf}].

This revised development evidently captures the same physical requirements for VEV as the original formulation. The difference between the original development and the constructions are the sets of test functions used to label states within the Hilbert space. Tempered functions are sufficient to determine the support of $\underline{W}$ since both the test functions and the Fourier transforms of test functions include functions of compact support. Negative energy support is excluded in construction of the physical states. Even though these constrained tempered functions do not include functions with compact spacetime support,\footnote{Elements of ${\cal B}$ have a form $i\dot{f}+\sqrt{m^2-\Delta}f$, $f\in {\cal A}$, and $f=0$ is the only function such that $i\dot{f}+\sqrt{m^2-\Delta}f$ vanishes in a domain [\ref{segal}]. ${\cal B}$ lacks real test functions and the natural association of equal-time fields with random processes does not apply when $\underline{W}$ is conditionally nonnegative.} the essential points of $\underline{W}$ are not modified in limiting the test functions used to construct the Hilbert space. A.IV is verified by assessing the essential points of the sums and differences of $\underline{W}$ required to be functionals for ${\cal A}$. Non-local functionals are excluded from ${\cal A}'$, the generalized functions for ${\cal A}$. $\underline{W}\in {\cal A}' \subset {\cal B}'$.

The Hilbert space ${\bf H}$ is constructed as the completion in the Wightman semi-norm of the linear vector space of equivalence classes in the quotient space ${\cal B}/{\cal N}$ with the null space ${\cal N}$ consisting of $\underline{h}\in {\cal B}$ with $\underline{W}(\underline{h}^*\,{\bf x}\,\underline{h})=0$.

\section{Wightman-functionals} 
For every free field $\underline{W_o}$, there is a family of non-trivial Wightman-functionals $\underline{W}$. One or more constituent elementary particles are included in the free field (\ref{twopoint}). In the construction, first $\underline{W}$ is defined as a linear functional to provide a semi-norm for ${\cal B}$ and then the demonstration is completed by verifying that $\underline{W}$ is continuous for ${\cal A}$. Poincar\'{e} covariance, spectral support, and locality are then demonstrated. Discussion includes description of the Hamiltonians, evaluation of the scattering amplitudes, and identification of the connected terms in $W_n$.


\subsection{A semi-norm} The unconditional nonnegativity of the free field functional, $\underline{W_o}(\underline{f}^* \, {\bf x}\, \underline{f})\geq 0$ [\ref{bogo},\ref{cook}], is exploited to achieve a non-trivial Wightman semi-norm for ${\cal B}$. Deformations of function sequences $\underline{f} \mapsto \underline{\rho}[\underline{f};\lambda_o]$ implement a semi-norm for ${\cal B}$. For the moment deferring the question of whether such a deformation results in a continuous functional, when the weights in the summation\begin{equation}\label{seminorm} \underline{W}(\underline{f}^* \, {\bf x}\, \underline{g}) := \int d\mu_o(\lambda_o)\; \underline{W_o}(\underline{\rho}[\underline{f};\lambda_o]^* \, {\bf x}\, \underline{\rho}[\underline{g};\lambda_o]),\end{equation}are nonnegative then\begin{equation}\label{norm} \| \underline{f} \|:= \sqrt{\underline{W}(\underline{f}^* \, {\bf x}\, \underline{f})}\end{equation}is a semi-norm for ${\cal B}$. The deformations $\underline{\rho}[\underline{f}]:=\underline{\rho}[\underline{f};\lambda_o]$ are determined below and $\lambda_o$ are parameters of the deformations. The deformation is not constrained to preserve algebraic properties of the operators realized in the free field base case.

{\em Digression to define the deformation.} The deformation derives from generators for the Wightman-functionals. The semi-norm is essentially a consequence of the Schur product and related theorems [\ref{horn}] that the Hadamard products of nonnegative matrices are nonnegative. The demonstration of a semi-norm is an elaboration on the development that a matrix is nonnegative, $\sum_{i,j}P_{ij}\overline{\alpha_i}\alpha_j \geq 0$, if and only if it has nonnegatively weighted factors $P_{ij} =\sum_k \overline{C_{ki}}C_{kj}$.

A generator for $\underline{W}$ is constructed from a symmetrized product of a generator ${\cal G}_o$ for the free field Wightman-functional $\underline{W_o}$, generators ${\cal G}_{n,m}$ for the higher order connected functions, and an energy ordering $\Theta_{k,n}$. Finite multinomials with generalized functions as coefficients suffice for the generators since ${\cal A}$ consists of terminating sequences. The components of the Wightman-functional are\begin{equation}\label{w-defn}\tilde{W}_n((\xi)_n):=\left({\ds \prod_{j=1}^n} \frac{\ds \partial\;}{\ds \partial\alpha_j}\right) \; {\ds \sum_{k=0}^n} \left( \frac{\ds {\bf S}[{\cal G}_{k,n-k}((\alpha,\xi)_n)\Theta_{k,n} {\cal G}_o((\alpha,\xi)_n)]}{\ds k!\, (n-k)!}\right)\end{equation}evaluated at $(\alpha)_n=0$.\begin{equation}\label{e-ord} \Theta_{k,n}:=\prod_{j=1}^k \theta(-E_j) \prod_{\ell=k+1}^n \theta(E_\ell).\end{equation}(\ref{w-defn}) is one of many forms that satisfy the revised Wightman conditions. This selected form exhibits a unique vacuum (is indecomposable [\ref{borchers}]) and reduces to a free field $\underline{W_o}$ when ${\cal G}_{n,m}=1$ as generalized functions for ${\cal B}$.

The product rule for differentiation results in the expansion [\ref{combin}]\begin{equation} \renewcommand{\arraystretch}{1.25} \begin{array}{l} {\ds \prod_{j=1}^{n+m}} \frac{\ds \partial\;}{\ds \partial\alpha_j} \; {\cal G}_{k,n+m-k}((\alpha,\xi)_{n\!+\!m}) {\cal G}_o((\alpha,\xi)_{n\!+\!m} )\\
 \qquad \qquad = {\ds \sum_{I_{\ell,n}}} {\ds \sum_{I_{\ell',m}}} \{ {\cal L}_{I'_{\ell,n}} {\cal L}_{I'_{\ell',m}} \; {\cal G}_{k,n+m-k}((\alpha,\xi)_{n\!+\!m})\} \{ {\cal L}_{I_{\ell,n}} {\cal L}_{I_{\ell',m}} \; {\cal G}_o((\alpha,\xi)_{n\!+\!m})\}\end{array} \label{inr} \end{equation}with\[{\cal L}_{I_{\ell,n}} :={\ds \prod_{j\in I_{\ell,n}}} \frac{\ds \partial\;}{\ds \partial\alpha_j}\]and derivatives apply only within the brackets $\{\ldots \}$. The set of integers $\{1,2,\ldots n\}$ is again designated $\chi_1^n$ and $I_{\ell,n}:=\{i_1,i_2,\ldots i_{\ell}\}$ is an element of the set of subsets of $\chi_1^n$ with exactly $\ell$ members. $I'_{\ell,n}$ is the complement of $I_{\ell,n}$ in $\chi_1^n$. $(p)_{I_{\ell,n}}:=p_{i_1},p_{i_2},\ldots p_{i_{\ell}}$ and $(p)_{I'_{\ell,n}}$ is similarly defined for $I'_{\ell,n}$. Summation over $I_{\ell,n}$ includes each of the binomial number ${_nC_\ell}$ subsets of $\chi_1^n$ with $\ell$ elements as well as a summation over $\ell=0,1,\ldots n$. Similarly for the subsets $I_{\ell',m}$ and $I'_{\ell',m}$ of $\chi_{n+1}^{n+m}:=\{n+1,n+2,\ldots n+m\}$.

The free field generator is defined by the two-point functional (\ref{twopoint}).\begin{equation} \renewcommand{\arraystretch}{1.25} \begin{array}{l} {\cal L}_{I_{\ell,n}} {\cal L}_{I_{\ell',m}} {\cal G}_o((\alpha,\xi)_{n\!+\!m}):=\tilde{W}_{o;\ell+\ell'}((\xi)_{I_{\ell,n}},(\xi)_{I_{\ell',m}})\\
 \qquad \qquad := \left\{ \renewcommand{\arraystretch}{1.5} \begin{array}{ll} {\ds \sum_{\mathit{pairs}}} s_{\pi_1\ldots \pi_{2k}} \tilde{\Delta}(\xi_{\pi_1},\xi_{\pi_2})\ldots \tilde{\Delta}(\xi_{\pi_{2k\!-\!1}},\xi_{\pi_{2k}}) & \quad \ell\!+\!\ell'=2k\\
 0 & \quad \ell\!+\!\ell'=2k\!+\!1 \end{array} \right. \end{array}\label{wodefn} \end{equation}when $(\alpha)_{n+m}=0$. $s_{\pi_1\ldots \pi_{2k}}$ is from (\ref{permutation}), $s_{12\ldots 2k}=1$ and the sum is over all $(2k)!/(2^k k!)$ pairs from $\chi_1^{2k}$ without regard to order. The indices of the two-point functionals are in ascending index order. Then (\ref{th1}) and (\ref{w-defn}) provide that\begin{equation}\label{th1-eval} \renewcommand{\arraystretch}{1.5} \begin{array}{l} \underline{W}(\underline{f}^*\, {\bf x}\, \underline{g})= {\ds \sum_{n,m} \int} (d\xi)_{n+m}\; ((D^T\cdot)_n \overline{\tilde{f}_n}((-\xi)_{n,1}))\;\tilde{g}_m((\xi)_{n+1,n+m}) {\ds \sum_{I_{\ell,n}}} {\ds \sum_{I_{\ell',m}}}\times\\
 \qquad \quad {\ds \sum_{k=0}^{n+m}}\; \frac{\ds {\bf S}[ {\cal L}_{I'_{\ell,n}} {\cal L}_{I'_{\ell',m}} {\cal G}_{k,n+m-k}((\alpha,\xi)_{n\!+\!m})\, \Theta_{k,n+m}\, \tilde{W}_{o;\ell+\ell'}((\xi)_{I_{\ell,n}},(\xi)_{I_{\ell',m}})]}{\ds k!\, (n+m-k)!}.\end{array}\end{equation}
 
When $f_n \in {\cal B}$,\begin{equation} \theta(-E_j)\, \tilde{f}_n((\xi)_n)=\theta(E_j)\, \tilde{f}_n^*((\xi)_n)=0 \label{epro}\end{equation}for all $j\in \chi_1^n$, and consequently\[{\bf S}[ \ldots \Theta_{k,n+m} \ldots]\overline{\tilde{f}_n}((-\xi)_{n,1}))\;\tilde{g}_m((\xi)_{n+1,n+m})=0\]unless $k=n$. Defining functionals from the generators of the higher order connected functions,\begin{equation} \tilde{V}_{\ell,\ell'}((\xi)_{I_{\ell,n}},(\xi)_{I_{\ell',m}}):= {\cal L}_{I_{\ell,n}} {\cal L}_{I_{\ell',m}} \; {\cal G}_{n,m}((\alpha,\xi)_{n\!+\!m})\label{Vdefn} \end{equation}when $(\alpha)_{n+m}=0$, the sums can be reorganized to write\begin{equation} \renewcommand{\arraystretch}{1.5} \begin{array}{l} \underline{W}(\underline{f}^*\, {\bf x}\, \underline{g})= 
{\ds \sum_{n,m} \int} (d\xi)_{n+m}\;\hat{\bf S}[\overline{\tilde{f}_n}((-\xi)_{n,1})]\;\hat{\bf S}[\tilde{g}_m((\xi)_{n+1,n+m})] \times\\
 \qquad \quad {\ds \sum_{I_{\ell,n}}} {\ds \sum_{I_{\ell',m}}} ((D\cdot)_{I'_{\ell,n}} \tilde{V}_{n-\ell,m-\ell'}((\xi)_{I'_{\ell,n}},(\xi)_{I'_{\ell',m}}))\, ((D\cdot)_{I_{\ell,n}} \tilde{W}_{o;\ell+\ell'}((\xi)_{I_{\ell,n}},(\xi)_{I_{\ell',m}})) \end{array} \label{H-G}\end{equation}when $\underline{f},\underline{g} \in {\cal B}$ and with\begin{equation}\label{shat} \hat{\bf S}[f_n] =\frac{1}{n!} {\bf S}[f_n].\end{equation}The generalized functions\[\tilde{V}_{\ell,\ell'}((\xi)_{I_{\ell,n}},(\xi)_{I_{\ell',m}}) :=\tilde{V}_{\ell,\ell'}((\xi)_{I_{\ell,n}},(\xi)_{I_{\ell',m}};n,I_{\ell,n},m,I_{\ell',m})\]include parameters $n,m$ and the sets of indices $I_{\ell,n},I_{\ell',m}$. (\ref{H-G}) follows from (\ref{th1-eval}) using (\ref{dualf}), the invariance of $(D^T\cdot)_n$ under permutations of $\chi_1^n$, the limited energy support of functions in ${\cal B}$, and that $I'_{\ell,n}$ has $n-\ell$ elements as the complement of $I_{\ell,n}$ in $\chi_1^n$. There is exactly one factor of either $\delta^+_j$ or $\delta^-_j$ for each argument $\xi_j$ resulting from (\ref{twopoint}) or $\tilde{V}_{n-\ell,m-\ell'}$. This is a condition on the generators ${\cal G}_{n,m}$ that is satisfied below in (\ref{genr}). (\ref{epro}) provides that the only terms that contribute to (\ref{H-G}) have a $\delta^-_j$ for each $j\in \chi_1^n$ and a $\delta^+_j$ for each $j\in \chi_{n+1}^{n+m}$. (\ref{H-G}) results from identifying that any transposition between arguments with one from $\chi_1^n$ and the other from $\chi_{n+1}^{n+m}$ vanishes due to (\ref{epro}), that permutations within $\chi_1^n$ and $\chi_{n+1}^{n+m}$ are proper subgroups of the symmetry group for $\chi_1^{n+m}$, and the signs (\ref{permutation}) of the permutations agree.
 
In ${\cal B}$, the contributing VEV coincide with the VEV for free fields when the interaction vanishes. When ${\cal G}_{n,m}=1$, $\tilde{V}_{0,0}=1$ and the other $\tilde{V}_{n,m}=0$. $\Theta_{k,n+m}\tilde{W}_{o;n+m}((\xi)_{n+m})$ contributes only when $n+m=2k$. From (\ref{wodefn}), only $k!$ of the $(2k)!/(2^kk!)$ terms in $\tilde{W}_{o;2k}((\xi)_{2k})$ contribute due to the energy ordering (\ref{e-ord}), those with two-point function factors with the index of the first argument from $\chi_1^k$ and the second from $\chi_{k+1}^{2k}$. From each contributing term, $(k!)^2$ of the $(2k)!$ permutations (\ref{permutation}) contribute in ${\cal B}$, the permutations limited to within $\chi_1^k$ and within $\chi_{k+1}^{2k}$. The permutations of each of the $k!$ contributing terms from $W_{o;2k}$ coincide resulting in ${\bf S}[\Theta_{k,2k} W_{o;2k}]=(k!)^2\, \Theta_{k,2k}W_{o;2k}=(k!)^2\, W_{o;2k}$ in evaluation of the sesquilinear functional (\ref{th1}) for elements from ${\cal B}$. Then (\ref{H-G}) reduces to the semi-norm for a free field for function sequences from ${\cal B}$ when ${\cal G}_{n,m}=1$.

The generators for the higher order connected functions ${\cal G}_{n,m}((\alpha,\xi)_{n\!+\!m})$ are Hadamard functions of the $M(p)_{\kappa_1 \kappa_2}$ from (\ref{twopoint}).\begin{equation} \renewcommand{\arraystretch}{1.25} \begin{array}{l} \ln\left({\cal G}_{n,m}((\alpha,\xi)_{n+m})\right) := {\ds \int} d\zeta_1 \; \overline{z_n((-\xi)_{n,1},M^*)}\,z_m((\xi)_{n\!+\!1,n\!+\!m},M)\times\\
 \qquad \qquad \exp({\ds \int} d\zeta_2 \; \overline{w_n((-\xi)_{n,1},DC)}\,w_m((\xi)_{n\!+\!1,n\!+\!m},C))  \end{array} \label{genr} \end{equation}with $DM=C^*C$ from (\ref{matcond}) and\begin{equation}\label{z-defn} \renewcommand{\arraystretch}{1.25} \begin{array}{rl} z_n((\xi)_{\eta\!+\!1,\eta\!+\!n},M) :=& \varsigma_n {\ds \prod_{\ell=\eta+1}^{\eta+n}} (a_\ell+ \lambda \alpha_{\ell} e^{-ip_{\ell} u } \delta^+_{\ell} \frac{\ds \partial\;}{\ds \partial\rho_{\ell}})\times\\
 &\qquad \qquad \exp({\ds \sum_{\eta<k<j}^{\eta+n}} \rho_k \rho_j \,U_n(p_k\!-\!p_j) M_{\kappa_k \kappa_j}(p_k\!-\!p_j) )\\
 w_n((\xi)_{\eta\!+\!1,\eta\!+\!n},C):=& {\ds \sum_{j=\eta+1}^{\eta+n}} \;e^{-(j-\eta)v\!-\!p_j(s'\!+\!s)} \rho_j C(s)_{\ell \kappa_{j}}. \end{array}\end{equation}The derivatives with respect to the $\rho_k$ apply to all factors and the generator is evaluated at $(\rho)_{n+m}=0$. ${\cal G}_{n,m}=1$ for $n,m=0,1$. The parameters of $z_n$ include the $(\alpha)_{\eta\!+\!1,\eta\!+\!n}$ and $\zeta_1:=\lambda,u$, and for $w_n$ the parameters include $\zeta_2:=s',s,v,\ell$. The $\varsigma_n$ are complex constants. The real constants\[a_\ell =\left\{\begin{array}{ll}0\qquad&\ell=1,2\\1 &\mbox{otherwise}\end{array}\right.\]result in a lowest contributing term in the generator ${\cal G}_{n,m}$ that is quartic in the $(\alpha)_{n+m}$ and removes the divergent two-point contribution that would result from extrapolating (\ref{genr}) to quadratic terms. $(\xi)_{n,1}$ indicates that the indices are in descending order, $n,n-1,\ldots 1$ implemented by mapping $\xi_j \mapsto \xi_{n+1-j}$ in the expression (\ref{z-defn}). $w_m((\xi)_{n\!+\!1,n\!+\!m},M)$ is independent of $n$ with $e^{-(j-n)v}$ evaluated for $j-n \in \{1,\ldots m\}$ although the $\rho_j,\xi_j$ are labeled with $j\in \chi_{n+1}^{n+m}$. Sums over $\eta<k<j$ indicate that the summation is over all integer $k$ and $j$ with $\eta < k < j \leq \eta+n$. The indicated summations are\[\renewcommand{\arraystretch}{1.5} \begin{array}{l} {\ds \int d\zeta_1 := \int d\sigma(\lambda)\int du}\\ {\ds \int d\zeta_2 := \int d\mu_u(s') \int d\mu_s(s)\int d\mu_\beta(v) \sum_{\ell=1}^{N_c}}. \end{array}\]These summations use nonnegative measure. $d\mu_s(s)$ and $d\mu_u(s)$ are  nonnegative, Lorentz invariant measures with support only for positive energies. These measures correspond with one-dimensional nonnegative tempered measures $d\mu_1(\lambda)$ [\ref{steinmann}] as\begin{equation}\label{bmat2} d\mu_s(s)=\left(a\delta(s)+\int d\mu_1(\lambda)\; \delta^+(s^2-\lambda)\right)\; ds.\end{equation}$d\sigma(\lambda)$ is a nonnegative measure with finite moments,\[c_n:=\int d\sigma(\lambda)\; \lambda^n.\]
 
The form (\ref{z-defn}) is selected to satisfy Poincar\'{e} covariance. (\ref{cond-D}) and (\ref{matcond}) provide that\begin{equation} \label{sum2}\renewcommand{\arraystretch}{1.5} \begin{array}{l} {\ds \int} d\zeta_2 \; \overline{w_n((-\xi)_{n,1},DC)}\,w_m((\xi)_{n\!+\!1,n\!+\!m}),C)={\ds \int d\mu_u(s')\int d\mu_s(s)\int d\mu_\beta(v) \sum_{\ell=1}^{N_c}}\times\\
\qquad \qquad \qquad \left({\ds \sum_{k=1}^n} e^{-kv\!+\!p_{n\!+\!1\!-\!k}(s\!+\!s')} \rho_k \overline{DC(s)}_{\ell \kappa_{n\!+\!1\!-\!k}} \right)\left({\ds \sum_{j=n+1}^{n+m}} e^{-\!(j\!-\!n)v\!-\!p_{j}(s\!+\!s')} \rho_j C(s)_{\ell \kappa_{j}}\right)\\
\qquad \qquad \quad = {\ds \sum_{k=1}^n \sum_{j=n+1}^{n+m}} \rho_k\rho_j\beta_{k+j-n}\Upsilon(-p_{n\!+\!1\!-\!k}\!+\!p_j) B_{\kappa_k \kappa_j}(-p_{n\!+\!1\!-\!k}\!+\!p_j)\end{array}\end{equation}with the definitions\begin{equation}\label{bmat} \renewcommand{\arraystretch}{1.25} \begin{array}{rl}B_{\kappa_k \kappa_j}(p) :=&{\ds \int} d\mu_s(s)\; M_{\kappa_k \kappa_j}(s)\; e^{-sp}\\ \Upsilon(p) :=&{\ds \int} d\mu_u(s)\; e^{-sp}\\
\beta_j:=&\int d\mu_\beta(v)\; e^{-jv}.\end{array}\end{equation}The functions $\Upsilon(p)$ and $U_n(p)$ from (\ref{z-defn}) are Lorentz scalars and multipliers of test functions, realized, for example, by polynomials and certain rational functions of $p^2$. There is no loss of generality in setting $\beta_2=1$ since a scale on the $\beta_j$ is redundant with scaling $\Upsilon(p)$.

In the evaluation of (\ref{H-G}), for every argument $\xi_k$ with $k\leq n$ from ${\cal G}_{n,m}((\alpha,\xi)_{n+m})$ there is a factor of $D$. In (\ref{genr}), there is either one factor of the complex conjugate of $DC(s)$ for every argument $\xi_k$ with $k\leq n$ or one factor of the complex conjugate of $M(p)^*$ for every pair of distinct arguments $\xi_k,\xi_{k'}$ with $k,k'\leq n$. (\ref{cond-D}), (\ref{matcond}) and (\ref{Mstar}) provide that $D \overline{DC(s)}=\overline{C(s)}$ and\[((D\cdot)_2 \overline{M(p)^*})= D\overline{DM(p)D^T}D^T =\overline{M(p)}.\]Consequently, the substitution\begin{equation} \renewcommand{\arraystretch}{1.5} \begin{array}{l} 
\ln\left(((D\cdot)_{I'_{\ell,n}} {\cal G}_{n,m}((\alpha,\xi)_{n+m}))\right) = {\ds \int} d\zeta_1 \; \overline{z_n((-\xi)_{n,1},M)}\,z_m((\xi)_{n\!+\!1,n\!+\!m},M)\times\\
 \qquad \qquad \exp({\ds \int} d\zeta_2 \; \overline{w_n((-\xi)_{n,1},C)}\,w_m((\xi)_{n\!+\!1,n\!+\!m}),C) \end{array} \label{genr1} \end{equation}as the generator of $((D\cdot)_n \tilde{V}_{n,m}((\xi)_{n+m})$ reproduces\begin{equation} \label{dtimes} ((D\cdot)_{I'_{\ell,n}} \tilde{V}_{n-\ell,m-\ell'}((\xi)_{I'_{\ell,n}},(\xi)_{I'_{\ell',m}}))= {\cal L}_{I'_{\ell,n}} {\cal L}_{I'_{\ell',m}} \; ((D\cdot)_{I'_{\ell,n}} {\cal G}_{n,m}((\alpha,\xi)_{n\!+\!m})) \end{equation}at $(\alpha)_{n+m}=0$.

The entire series for the exponential function\[\exp({\ds \int} ds\; \overline{u_n(s)} u_m(s))={\ds \sum_\nu} \frac{\ds 1}{\ds \nu!} {\ds \int}(ds)_\nu\; (\overline{u_n(s)})_\nu (u_m(s))_\nu\]puts the result into the desired form. The repetition of variables now includes a summation over a set parameters $\zeta_1$ or $\zeta_2$ for each factor. (\ref{H-G}), (\ref{genr1}) and (\ref{dtimes}) express (\ref{th1}) as\begin{equation} \label{result0} \renewcommand{\arraystretch}{1.25} \begin{array}{l} \langle \underline{f}|\underline{g}\rangle =\left. {\ds \sum_{n,m} \sum_{I_{\ell,n}} {\cal L}_{I'_{\ell,n}} \sum_{I_{\ell',m}} {\cal L}_{I'_{\ell',m}} \int} (d\xi)_{n+m} \;{\ds \sum_{\nu_1}} \frac{\ds 1}{\ds \nu_1!} {\ds \int}(d\zeta_1)_{\nu_1} \right( \overline{z_n((-\xi)_{I'_{\ell,n}\downarrow},M)}\times \\
 \qquad \left. z_m((\xi)_{I'_{\ell',m}},M) \;{\ds \sum_{\nu_2}} \frac{\ds 1}{\ds \nu_2!} {\ds \int}(d\zeta_2)_{\nu_2} \left( \overline{w_n((-\xi)_{I'_{\ell,n}\downarrow},C)}w_m((\xi)_{I'_{\ell',m}},C) \right)_{\nu_2}\right)_{\nu_1}\times\\
 \qquad ((D\cdot)_{I_{\ell,n}} \tilde{W}_{o;\ell+\ell'}((\xi)_{I_{\ell,n}},(\xi)_{I_{\ell',m}}))\,\hat{\bf S}[\overline{\tilde{f}_n}((-\xi)_{n,1})]\,\hat{\bf S}[\tilde{g}_m((\xi)_{n+1,n+m})]. \end{array} \end{equation}This is evaluated at $(\alpha,\rho)_{n+m}=0$ and the summations are absolutely convergent. The $\zeta_1$ in $z_n,z_m$ and $\zeta_2$ in $w_n,w_m$ are indexed appropriately for each indicated repetition following (\ref{recursion}). $(\xi)_{I_{\ell,n}\downarrow}$ indicates that the arguments are in descending index order. Since ${\cal B}$ consists of terminating sequences, finite expansions in the $\alpha_k$ and $\rho_k$ suffice.

Reordering the summations and organizing (\ref{result0}) defines the form sought, (\ref{seminorm}).\[\renewcommand{\arraystretch}{1.25} \begin{array}{l} \langle \underline{f}|\underline{g}\rangle = {\ds \sum_{\nu_1} \frac{1}{\nu_1!} (\int d\zeta_1\;\sum_{\nu_2} \frac{1}{\nu_2!}(\int d\zeta_2)_{\nu_2})_{\nu_1}} {\ds \sum_{\ell,\ell'} \sum_{n,m} \sum_{I_{\ell,n}|_{\ell}}\sum_{I_{\ell',m}|_{\ell'}}} {\ds \int} (d\xi)_{I_{\ell,n}} (d\xi)_{I_{\ell',m}}\times\\

 \qquad ((D\cdot)_{I_{\ell,n}} \tilde{W}_{o;\ell+\ell'}((\xi)_{I_{\ell,n}},(\xi)_{I_{\ell',m}})) \times\\
 \qquad {\cal L}_{I'_{\ell,n}} {\ds \int} (d\xi)_{I'_{\ell,n}} \left( \overline{z_n((-\xi)_{I'_{\ell,n}\downarrow},M)} \left(\overline{w_n((-\xi)_{I'_{\ell,n}\downarrow},C)} \right)_{\nu_2}\right)_{\nu_1} \hat{\bf S}[\overline{\tilde{f}_n}((-\xi)_{n,1})]\times\\
 \qquad {\cal L}_{I'_{\ell',m}} {\ds \int} (d\xi)_{I'_{\ell',m}} \left( z_m((\xi)_{I'_{\ell',m}},M) \left(w_m((\xi)_{I'_{\ell',m}},C) \right)_{\nu_2}\right)_{\nu_1}\hat{\bf S}[\tilde{g}_m((\xi)_{n+1,n+m})]
\end{array} \]at $(\alpha,\rho)_n=0$. $\sum_{I_{\ell,n}|_{\ell}}$ denotes the sum over those subsets $I_{\ell,n}\subset \chi_1^n$ with exactly $\ell$ members. The deformation $\underline{\rho}[\underline{f}]$ can be defined to put this in the form of (\ref{seminorm}). With the arguments of the generalized functions $\tilde{W}_{o;\ell+\ell'}((\xi)_{\ell+\ell'})$ relabeled appropriately term by term as $(\xi)_{I_{\ell,n}},(\xi)_{I_{\ell',m}}$, the $\ell$th component of $\underline{\rho}[\underline{f}]$ is\begin{equation} \label{result1}\rho_{\ell}[\underline{f}] ={\ds \sum_n \sum_{I_{\ell,n}|_{\ell}} {\cal L}_{I'_{\ell,n}} \int} (d\xi)_{I'_{\ell,n}} \left( z_n((\xi)_{I'_{\ell,n}},M) \left( w_n((\xi)_{I'_{\ell,n}},C) \right)_{\nu_2}\right)_{\nu_1}\hat{\bf S}[\tilde{f}_n((\xi)_n)]\end{equation}with $(\alpha,\rho)_n=0$. The $\ell$ arguments of $\rho_{\ell}[\underline{f}]$ are $(\xi)_{I_{\ell,n}}$ relabeled by term. The parameters of the deformation $\underline{\rho}[\underline{f}]$ are\[\lambda_o=\nu_1,(\nu_2)_{\nu_1},(\zeta_1)_{\nu_1},((\zeta_2)_{\nu_2})_{\nu_1}\]and\[\int d\mu_o(\lambda_o) = \sum_{\nu_1} \frac{1}{\nu_1!} (\int d\zeta_1\;\sum_{\nu_2} \frac{1}{\nu_2!}(\int d\zeta_2)_{\nu_2})_{\nu_1}.\](\ref{seminorm}) results from (\ref{dualf}), (\ref{sdual}) and relabeling summations $\xi_i\mapsto -\xi_{n+1-i}$ for $i\in I'_{\ell,n}$. {\em End digression.}

The result of this paper can be stated as:

\begin{quote}{\em Theorem.} (\ref{seminorm}) and (\ref{result1}) with the attendant definitions are sufficient conditions for a QFT to satisfy A.1-A.V.\end{quote}

(\ref{result1}) substituted in (\ref{seminorm}) defines (\ref{th1}) as a semi-norm for ${\cal B}$ when $\underline{W}$ is a functional. The Jost-Schroer and related theorems [\ref{bogo},\ref{mund}] provides that this semi-norm can not be extended to ${\cal A}$ and preserve satisfaction of the Poincar\'{e} covariance, locality, and spectral support conditions with the exception of free fields. Extension of the semi-norm to ${\cal A}$ would produce the contradiction to the Jost-Schroer theorem of a QFT with symmetric Hilbert space field operators exhibiting interaction and with the two-point function of a free field. 

These constructions are described by: a two-point function (\ref{twopoint}) that determines the constituent elementary particles; coefficients $c_n$ that are the moments of a nonnegative measure $d\sigma(\lambda)$; complex constants $\varsigma_n$; Lorentz invariant functions $U_n(p),\Upsilon(p)$ that are multipliers of tempered functions; coefficients $\beta_j$ that are Laplace transforms (\ref{bmat}) of a nonnegative measure $d\mu_{\beta}(v)$; and a nonnegative, Lorentz invariant measure $d\mu_s(s)$ (\ref{bmat2}). Like free fields, the constructed $W_n$ vanish unless $n$ is even, although a Wightman semi-norm with odd $n>3$ contributing is readily demonstrated for scalar models and addition of a character [\ref{heger}] achieves a vacuum polarization and trivial odd functions. Several methods [\ref{yngvason}] generate additional $\underline{W}$ given realizations of A.1-A.V. 

\subsection{Continuity}
$\underline{W}$ is a continuous linear functional [\ref{gel1}]. In the evaluation of (\ref{H-G}), the generalized function is a sum of terms that are products of functionals $\tilde{W}_{o;n}((\xi)_n)$ and $\tilde{V}_{n,m}((\xi)_{n+m})$ that share no arguments. The numbers of terms are finite for terminating sequences $\underline{f}$. $\tilde{W}_{o;n}((\xi)_n)$ is the free field VEV and well defined. $\tilde{V}_{n,m}((\xi)_{n+m})$ is also a finite sum of terms that are products of factors that share no arguments and from (\ref{genr}), (\ref{z-defn}) and (\ref{sum2}), the factors are of the form\begin{equation}\label{expand-f}\renewcommand{\arraystretch}{1.25} \begin{array}{l}
T_n((\xi)_{I_n}) = (2\pi)^d c_n\;\delta(p_{i_1}+\ldots p_{i_n}) \; \left( \delta^-_{i_1}\delta^-_{i_2} \overline{M}_{\kappa_{i_1},\kappa_{i_2}}(p_{i_1}\!-\!p_{i_2}) \right.\ldots \times\\ 
 \qquad \qquad \left. \delta^-_{i_{k-1}}\delta^+_{i_k} \beta_{1\!-\!i_{k\!-\!1}\!+\!i_k}\, B_{\kappa_{i_{k-1}} \kappa_{i_k}}(-p_{i_{k-1}}\!+\!p_{i_k})\ldots \delta^+_{i_{n-1}}\delta^+_{i_n} M_{\kappa_{i_{n-1}} \kappa_{i_n}}(p_{i_{n-1}}-p_{i_n})\right). \end{array} \end{equation}The $U_n(p)$ and $\Upsilon(p)$ are multipliers [\ref{gel2}] and need not be considered for continuity. $I_n=\{i_1,\ldots i_n\}$ is a set of distinct indices and $n\geq 4$. The factors in $T_n$ are any combination of $\delta^- \delta^- M$, $\delta^+ \delta^+ M$ and $\delta^- \delta^+ B$ that from (\ref{z-defn}) result in at least two factors of $\delta^-$ and two factors of $\delta^+$. $T_n$ vanishes for odd $n$. When $M(p)$ and $B(p)$ are multipliers of tempered test functions, the only demonstration required for continuity is that the product of the delta function that conserves energy-momentum with the $n$ mass shell delta functions defines a generalized function. This demonstration, in Appendix A, succeeds in three or more dimensions $d$ for finite masses $m_\kappa>0$.

The $M(p)$ presented in Appendix B are multipliers since the elements are multinomials in the components of $p$. $\delta(s)$ in (\ref{bmat2}) is the only finite Lorentz invariant measure and results in a trivial $B(p)$ from (\ref{bmat}). More generally, $B(p)$ is singular at $p^2=0$ due to the divergence of $\int d\mu_s(s)$. However, this singularity is excluded from the support of $\underline{W}$ in the form (\ref{genr}). Every factor $B(-p_i+p_j)$ is evaluated for $p_i,p_j$ on mass shells with $i\in \chi_1^n$, $j\in \chi_{n+1}^{n+m}$ and energies, $E_i\leq -m_{\kappa_i}$ and $E_j\geq m_{\kappa_j}$. The argument of every $B(-p_i+p_j)$ has positive energy and\begin{equation}\label{regularity} (p_i\!-\!p_j)^2\geq (m_{\kappa_i}+m_{\kappa_j})^2 >0.\end{equation}

A bound on $d\mu_s(s)$ ensures that the elements of $B(p)$ are multipliers. There is a Lorentz transformation to the center-of-mass frame for the time-like $p_j-p_i$. In this frame ${\bf p}_i={\bf p}_j$ and neglecting a finite contribution should $a\neq 0$ in (\ref{bmat2}), (\ref{bmat}) results in\begin{equation} \renewcommand{\arraystretch}{2} \begin{array}{rl} {\cal P}(\frac{d\;}{dp_i},\frac{d\;}{dp_j}) B_{\kappa_i \kappa_j}(-p_i\!+\!p_j)&={\ds \int} d\mu_s(s)\; {\cal P}(s,-s)M_{\kappa_i \kappa_j}(s)\; e^{-s_{(0)} (-E_i\!+\!E_j)}\\
 &={\ds \int} d\mu_1(\lambda)\; {\ds \int}ds \; \delta^+(s^2-\lambda)\; {\cal M}_{\kappa_i \kappa_j}(s)\; e^{-s_{(0)} (-E_i\!+\!E_j)}\\
 &={\ds \int} d\mu_1(\lambda)\; {\ds \int} {\ds \frac{d{\bf s}}{2\omega_{\lambda}}} \;{\cal M}_{\kappa_i \kappa_j}(\omega_{\lambda},{\bf s})\; e^{-\omega_{\lambda} (-E_i\!+\!E_j)}
 \end{array}\label{brep}\end{equation}with ${\cal M}_{\kappa_i \kappa_j}(s)={\cal P}(s,-s)M_{\kappa_i \kappa_j}(s)$, ${\cal P}(x,y)$ a multinomial and $\omega_{\lambda}^2 = \lambda +{\bf s}^2$. Then there are positive integers $N_1$ and $N_2$ such that\begin{equation}\label{bound-m} \sup_{{\bf s},\, \lambda>0} (1+\lambda^{N_1})^{-1} (1+{\bf s}^{2{N_2}})^{-1} \left| {\cal M}_{\kappa_i \kappa_j}(\omega_{\lambda},{\bf s}) \right|< C_1.\end{equation}The inequalities $\sqrt{2(\lambda+{\bf s}^2)} \geq \sqrt{\lambda}+\sqrt{{\bf s}^2}$ and $-E_i+E_j\geq m_{\kappa_i}+m_{\kappa_j}$ provide a bound\begin{equation}| {\cal P}({\textstyle \frac{d\;}{dp_i},\frac{d\;}{dp_j}}) B_{\kappa_i \kappa_j}(-p_i\!+\!p_j)| < C_2 {\ds \int} d\mu_1(\lambda)\;(1+\lambda^{N_1}) e^{-(m_{\kappa_i}+m_{\kappa_j})\sqrt{\lambda/2}}\end{equation}with\[C_2 = C_1 {\ds \int} {\ds \frac{d{\bf s}}{2\omega_{\lambda}}} \;(1+{\bf s}^{2{N_2}}) \; e^{-(m_{\kappa_i}\!+\!m_{\kappa_j})\sqrt{{\bf s}^2/2}}.\]The summation for $C_2$ is given by the volume of the unit sphere in $d-1$ dimensions and gamma functions. Then if\begin{equation} {\ds \int} d\mu_1(\lambda)\;(1+\lambda^{N_1}) e^{-(m_{\kappa_i}+m_{\kappa_j})\sqrt{\lambda/2}} < C_3, \label{convcrit} \end{equation}$B(p)$ is a multiplier since similar bounds also apply to $p^{2k}B(p)$. The Poincar\'{e} invariance of amplitudes (\ref{innerp}) then provides that $B(p)$ are multipliers when (\ref{convcrit}) holds.

Then, sufficient conditions to define $\underline{W}$ as a continuous linear functional for ${\cal A}$ are: $M(p)$ a multinomial in the energy-momentum components, masses are finite, $d\geq 3$, and (\ref{convcrit}).

\subsection{Poincar\'{e} covariance}

The constructed $\underline{W}$ satisfy A.II. Orthochronous Poincar\'{e} transformations preserve the sense of energies and consequently ${\cal B}$ is closed under orthochronous Poincar\'{e} transformations. Invariance of amplitudes (\ref{innerp}) follows from (\ref{matcond2}) and (\ref{p-cov}). There is exactly one factor of $S(A)$ for each argument $\xi_k$ and consequently two factors of $S(A)$ for every factor of $M(p)$ or $B(p)$ in $W_n$ (from (\ref{w-defn}), (\ref{genr}), (\ref{sum2}) and (\ref{bmat})).  Lorentz covariance (\ref{p-cov}) follows from $((S(A)\cdot)_2 M(p))=S(A)M(p)S(A)^T$ since the $M(p),D,S(A)$ presented in Appendix B satisfy (\ref{matcond2}) and $B(p)$ is covariant for the same Lorentz transformation as $M(p)$.\begin{equation} \renewcommand{\arraystretch}{1.5} \begin{array}{rl}
S(\Lambda)B(p)S(\Lambda)^T=&{\ds \int} d\mu_s(s)\; M(\Lambda^{-1}s)\; {\ds e^{-sp}}\\
=&{\ds \int} d\mu_s(s')\; M(s')\; {\ds e^{-s'(\Lambda^{-1}p)}}\\
=&B(\Lambda^{-1}p)\end{array} \end{equation}using the Lorentz invariance of the measure $d\mu_s(s)$, the substitution $s'=\Lambda^{-1}s$, $ps= p^Tgs$ and $\Lambda^Tg\Lambda =g$ for Lorentz transformations and the Minkowski signature $g$. Whether from (\ref{twopoint}) or (\ref{expand-f}), the $\delta(p_{i_1}+\ldots p_{i_n})$ and $\delta^\pm_{i_k}$ are Lorentz scalars for orthochronous transformations. 

Translation invariance is verified by examining (\ref{inr}) and (\ref{genr}). The product of factors in $W_n$ includes\begin{equation}\label{psupport}\delta(p_{i_1}+\ldots p_{i_\ell})\ldots \delta(p_{i_i}+\ldots p_{i_n})\end{equation}where each $p_k$ appears exactly once in the set of arguments. Then $p_1+p_2+\ldots p_n=0$ that results in translation invariance.

\subsection{Spectral support}

For a selection of $(\kappa)_n$, the support of $\tilde{W}_n((\xi)_n)$ considered as a generalized function for test functions $f_k^* g_{n-k}$ with $f_k,g_{n-k} \in {\cal B}$ is given by (\ref{twopoint}) and (\ref{expand-f}). As a consequence of (\ref{psupport}), the mass shell deltas and the support of $f_k^* g_{n-k}$, the essential points of each term within $\tilde{W}_n((\xi)_n)$ are contained within the set ${\cal Q}_{k,n}$ of points $(p)_n$ that cover the support of\[\delta(p_1+p_2+\ldots p_n) \prod_{j=1}^k \delta^-_j \prod_{j'=k+1}^n \delta^+_{j'}\]with the $k$ appropriate for the term. Then $p_1 +p_2 \ldots +p_n =0$ explicitly, and $p_n \in \bar{V}^+$, $p_{n-1} +p_n \in \bar{V}^+$, through $p_{k+1} +p_{k+2} \ldots +p_n \in \bar{V}^+$ due to closure of the closed forward $\bar{V}^+$ cone under addition of elements. From translation invariance,\[ \sum_{j=\ell}^n p_j = -\sum_{j=1}^{\ell-1} p_j \in \bar{V}^+\]for $\ell \leq k$ since each $p_j \in \bar{V}^-$ for $j\leq k$. The backward cone is also closed under addition and the negative of a vector in the backward cone is in the forward cone. Then every point $(p)_n \in {\cal Q}_{k,n}$ satisfies $p_n$, $p_{n-1}+p_n$ through $p_2+ \ldots +p_{n-1}+p_n \in \bar{V}^+$ and A.III is satisfied [\ref{pct},\ref{borchers}].

Since the constructed $\underline{W}$ are solutions of the Klein-Gordon equation and satisfy A.III, evolution with time is analogous to the time evolution of a free field.\begin{equation}\label{hamil} \renewcommand{\arraystretch}{2} \begin{array}{l} \langle \underline{f}| U(t) \underline{g} \rangle = {\ds \sum_{n,m} \;\sum_{\kappa_1=1}^{N_c}\ldots \sum_{\kappa_{n+m}=1}^{N_c} \int} (dx)_{n+m}\; ((D\cdot)_n W_{n\!+\!m}((x)_{n\!+\!m}))_{\kappa_1\ldots \kappa_{n\!+\!m}}\times \\
 \qquad \qquad \qquad \qquad \overline{f_n}((x)_{n,1})_{\kappa_n \ldots \kappa_1} \; g_m((x_{(0)}-t,{\bf x})_{n+1,n+m})_{\kappa_{n+1} \ldots \kappa_{n+m}}\\
\qquad = {\ds \sum_{n,m} \int} d(\xi)_{n+m}\;((D\cdot)_n \tilde{W}_{n+m}((\xi)_{n+m})) \overline{\tilde{f}_n((-\xi)_{n,1})}{\ds \prod_{k=n+1}^{n+m}} e^{-i\omega_k t}\tilde{g}_m((\xi)_{n+1,n+m}) \end{array}\end{equation}for $\underline{f}, \underline{g} \in {\cal B}$. The association with a classical limit is the interaction represented in transition amplitudes and not the form of the Hamiltonian.

\subsection{Locality}

Locality, A.IV, requires that the Wightman functions are symmetric under transpositions of adjacent space-like separated arguments.\begin{equation}\label{local}W_n(\ldots x_{\ell}, x_{\ell\!+\!1} \ldots)_{\ldots \kappa_{\ell}\kappa_{\ell\!+\!1}\ldots} = \sigma_{\kappa_{\ell},\kappa_{\ell\!+\!1}} W_n(\ldots x_{\ell\!+\!1}, 
x_{\ell} \ldots )_{\ldots \kappa_{\ell\!+\!1}\kappa_{\ell}\ldots}\end{equation}when $(x_{\ell}- x_{\ell\!+\!1})^2<0$ and with $\sigma_{\kappa_{\ell},\kappa_{\ell\!+\!1}}$ defined in (\ref{signs-perm}). The revised axioms A.1-A.V relax the conventional constraint of locality, enabling locality to be satisfied by symmetrization of the VEV as generalized functions on ${\cal A}$. The indicated symmetry is an immediate consequence of (\ref{permutation}) and (\ref{w-defn}). There are no localizable states in this development when interaction is exhibited (when (\ref{genr}) contributes). All states are labeled by $\underline{f}\in {\cal B}$ that consists of complex functions that do not vanish in any spatial domain, but the VEV are functionals on the larger set of test functions ${\cal A}$ that includes functions of compact spacetime support. On ${\cal B}\subset {\cal A}$, the difference of local functionals that satisfy the spectral support condition on ${\cal B}$ and non-local functionals that satisfy the spectral support condition in ${\cal A}$ vanish. Many of the terms added in the symmetrization do not contribute in ${\cal B}$. This implementation of locality also ensures symmetry of the states (\ref{H-G}).

\subsection{Interaction and asymptotic states}

The constructed VEV are singular on mass shells and exhibit non-trivial interaction when (\ref{genr}) contributes.

In free field models, LSZ states evaluate the VEV for creation operators and more generally these LSZ states are used to evaluate scattering amplitudes. For a single argument and with the LSZ state labeled $\ell(t;x)_\kappa$,\[ \renewcommand{\arraystretch}{2} \begin{array}{rl} i \langle \underline{h}| {\ds \int} d{\bf x}\; \hat{f}(x)_\kappa \stackrel{\leftrightarrow}{\partial}_o \Phi(x)_\kappa \;\underline{g} \rangle :=& \langle \underline{h}| {\ds \int} dp\; (\omega+E) e^{i(\omega-E)t} \tilde{f}({\bf p})_\kappa \;\tilde{\Phi}(p)_\kappa \;\underline{g} \rangle\\
=& \langle \underline{h}| U(-t)\Phi(\ell(t)_{\kappa})_\kappa U(-t)^{-1}\;\underline{g} \rangle\end{array}\]with $\tilde{f}({\bf p})$ a tempered test function, $U(t)$ is unitary time translation\[U(t)\Phi(x)U(t)^{-1}=\Phi(x_{(0)}-t,{\bf x})\]and\begin{equation}\label{skg}\hat{f}(x)_\kappa =\frac{1}{(2\pi)^{(d-2)/2}}\int d{\bf p}\; e^{i\omega t}e^{-i{\bf p}\cdot {\bf x}}\tilde{f}({\bf p})_\kappa\end{equation}is a smooth solution of the Klein-Gordon equation. The Fourier transform of an $\ell(t;x)_{\kappa} \in {\cal B}$ suitable for a plane wave limit is\begin{equation} \label{testf} \tilde{\ell}(t;p)_{\kappa}= (\omega+E)e^{i\omega t}\; \delta_L ({\bf p} -{\bf q})\;u_{\kappa}. \end{equation}with $\tilde{f}({\bf p})_\kappa =\delta_L({\bf p})u_\kappa$ and $\delta_L({\bf p})$ a test function delta sequence. The polarization and particle species are described by $u_{\kappa}$ and support is concentrated near the momentum ${\bf q}$ in the plane wave limit as $L \rightarrow \infty$. Plane wave ``in'' states are\[\lim_{\stackrel{L \rightarrow \infty}{t\rightarrow -\infty}} |U(-t)\tilde{\ell}_n(t) \rangle \rightarrow |(q,u)_n^{\mathit{in}} \rangle \]and ``out'' states are the $t\rightarrow \infty$ limits recognizing that the limits of transition probabilities may remain finite even when limits are not states within the Hilbert space. Evaluation of the scattering amplitudes below uses the pointwise nonnegative\[ \delta_L ({\bf p} -{\bf q})= \left(\frac{L}{\sqrt{\pi}} \right)^{d-1} \; e^{-L^2 ({\bf p} -{\bf q})^2}>0. \]
 
With energy-momentum support limited to mass shells,\[\renewcommand{\arraystretch}{1.5} \begin{array}{rl} i \langle \underline{h}| {\ds \int} d{\bf x}\; \hat{f}(x) \stackrel{\leftrightarrow}{\partial}_o \Phi(x)_\kappa \;\underline{g} \rangle &= \langle \underline{h}| {\ds \int dp}\; (\omega+E) e^{i(\omega-E)t} \tilde{f}({\bf p})\;\tilde{\Phi}(p)_\kappa \;\underline{g} \rangle\\
 &= \langle \underline{h}| {\ds \int dp}\; 2\omega \tilde{f}({\bf p})\;\tilde{\Phi}(p)_\kappa \;\underline{g} \rangle \end{array}\]is independent of time for smooth solutions of the Klein-Gordon equation $\hat{f}(x)$, (\ref{skg}). In these cases,\[\langle \underline{h}| {\ds \int} d{\bf x}\; \hat{f}(x) \stackrel{\leftrightarrow}{\partial}_o \Phi(x)_\kappa \;\underline{g} \rangle \neq \langle \underline{h}| {\ds \int} d{\bf x}\; \hat{f}(x) \stackrel{\leftrightarrow}{\partial}_o \Phi_o(x)_{\kappa} \;\underline{g} \rangle\]at asymptotic times with $\Phi_o(x)_{\kappa}$ a free field operator. For particular (time-dependent) states and the constructions of physical interest, the $n>2$ connected functions contribute for all times.


Transition amplitudes are\begin{equation} \label{scatt-def} \renewcommand{\arraystretch}{1.5} \begin{array}{rl} \langle (q,u)_k^{\mathit{out}} |(q,u)_{k\!+\!1,n}^{\mathit{in}} \rangle &= {\ds \lim}\;W_n(\ell_k(0)^* \,{\bf x}\,\ell_{n-k}(0))\\
 &= {\ds \sum_{\kappa_1=1}^{N_c} \ldots \sum_{\kappa_n=1}^{N_c}} \bar{u}_{\kappa_1} \ldots \bar{u}_{\kappa_k} u_{\kappa_{k+1}} \ldots u_{\kappa_n} \langle (q,\kappa)_k^{\mathit{out}} |(q,\kappa)_{k\!+\!1,n}^{\mathit{in}} \rangle \end{array}\end{equation}as $t,L \rightarrow \infty$. The probability of observing the state described as $u$ given a pure state prepared as $v$ is the magnitude of the projection of $v$ on $u$, $|\langle u | v\rangle |^2/(\langle u | u\rangle \langle v | v\rangle)$ [\ref{vonN}]. Only the connected function components ${^CW}_n$ contribute to the non-forward, plane wave limits. These connected function components ${^CW}_n$ are identified as the connected contribution to $W_n$ from (\ref{w-defn}). The connected terms vanish when any nonempty proper subset of arguments is separated from the remaining arguments by $\rho a$ with $a^2<0$ as the real number $\rho$ grows without bound. Using (\ref{perm-sub}) and (\ref{genr1}), evaluation of the $\delta_i^\pm$ and the Gaussian quadratures in (\ref{H-G}) result in\begin{equation} \label{result9} \renewcommand{\arraystretch}{1.25} \begin{array}{l}
\langle (p,\kappa)_k^{\mathit{out}} |(p,\kappa)_{k+1,n}^{\mathit{in}} \rangle =(2\pi)^d c_n \;\frac{\ds \overline{\varsigma_k}\, \varsigma_{n\!-\!k}}{\ds k!(n\!-\!k)!} \, \delta({\ds \sum_{\ell=1}^n} s_\ell p_\ell)  \left({\ds \prod_{\ell'=1}^n} \frac{\ds \partial\;}{\ds \partial\rho_{\ell'}} \right){\bf S}_{\chi_1^k}[{\bf S}_{\chi_{k+1}^n}[ \times\\
 \quad \exp\left({\ds \sum_{i<j}^k} \rho_i \rho_j \,\overline{U_k(p_i\!-\!p_j) M_{\kappa_i \kappa_j}(p_i\!-\!p_j)}+{\ds \sum_{k<i'<j'}^n} \rho_{i'} \rho_{j'} \,U_{n\!-\!k}(p_{i'}\!-\!p_{j'}) M_{\kappa_{i'} \kappa_{j'}}(p_{i'}\!-\!p_{j'}) \right) \times\\
 \quad \exp \left( {\ds \sum_{i"=1}^k \sum_{j"=k+1}^n} \rho_{i"} \rho_{j"} \beta_{i"\!+\!j"\!-\!k}\Upsilon(p_{i"}\!+\!p_{j"}) (DB)_{\kappa_{i"} \kappa_{j"}}(p_{i"}\!+\!p_{j"}) \right)]]\end{array} \end{equation}
evaluated with $(\rho)_n=0$, each $E_i = \omega_i$, and $s_\ell=-1$ for $\ell>k$ and $s_\ell=1$ otherwise. This result applies for the non-forward, plane wave limit with corrections of order $L^{-1}$ from approximations of the relatively slowly varying functions as constant within the momentum summations. The mean value theorem for integration, the nonnegativity of the delta sequences used in the LSZ functions, and continuity of $M(p)$, $B(p)$, $\Upsilon(p)$, $U_k(p)$, $\omega_j$ and a continuous regularization for the energy conserving delta justifies the approximation. In (\ref{result9}), the range of the summation over each component of $u$ producing the energy conserving delta is limited to $-1/\epsilon$ through $1/\epsilon$ and then the constant $\epsilon$ controls the approximation.\[\int_{-1/\epsilon}^{1/\epsilon}du_{(\ell)}\; e^{-iq_{(\ell)}u_{(\ell)}}=\frac{2\sin(q_{(\ell)}/\epsilon)}{q_{(\ell)}}\rightarrow 2\pi \delta(q_{(\ell)})\]as $\epsilon \rightarrow 0$ for each component of $q$. The forward contributions provide the state norms $\langle u|u \rangle$ [\ref{iqf}].

\subsection{Connected functions}

As generalized functions for ${\cal B}$, (\ref{H-G}) applies and the VEV have a connected term that vanishes when any nonempty proper subset of arguments are separated without bound in a space-like direction. This connected term results when all derivatives apply to the generator ${\cal G}_{n,m}$ in (\ref{w-defn}). From $\ln ({\cal G}_{n,m}{\cal G}_o)=\ln ({\cal G}_{n,m}) +\ln ({\cal G}_o)$, the connected component is\begin{equation}\label{cluster-cond}\renewcommand{\arraystretch}{1.25} \begin{array}{rl}  {\ds \lim_{r\rightarrow \infty}} {^CW}_n(f_k^* \,(ra,1) g_{n-k})&={\ds \lim_{r\rightarrow \infty} \int} (d\xi)_n\;\tilde{f}_k^*((\xi)_k)\;\tilde{g}_{n\!-\!k}((\xi)_{k\!+\!1,n}) \left(e^{-irap}\right)_{k\!+\!1,n}\times\\
 &\qquad \qquad \qquad \left( {\ds \prod_{k=1}^n} \frac{\ds \partial \;}{\ds \partial \alpha_k}\right) \ln( {\cal G}_{k,n-k}((\alpha,\xi)_n)) \rightarrow 0\end{array}\end{equation}as $|r|\rightarrow \infty$ for $n>3$ and $(\alpha)_{n\!+\!m}=0$ and in the notation of A.II. The vanishing of (\ref{cluster-cond}) is implied by\begin{equation}\renewcommand{\arraystretch}{2} \begin{array}{l} {\ds \lim_{r\rightarrow \infty} \int} (dq)_n \; \delta(q_n)\;\overline{\tilde{f}_k()} \tilde{g}_{n\!-\!k}() \,e^{-iraq_k} \; {\ds \prod_{\ell=1}^n \left(\frac{\ds \partial\;}{\ds \partial\rho_\ell} \right)} \times\\
 \qquad \left(\exp({\ds \sum_{i<j}^k} \rho_i \rho_j \, \delta^-_i \delta^-_j\overline{M^*_{\kappa_i \kappa_j}()} )\exp({\ds \sum_{k < i'<j'}^n } \rho_{i'} \rho_{j'} \, \delta^+_{i'} \delta^+_{j'} M_{\kappa_{i'} \kappa_{j'}}() )\right. \times \\
 \qquad \quad \left. \exp({\ds \sum_{i"=1}^k \sum_{j"=k+1}^n} \rho_{i"} \rho_{j"} \beta_{i"+j"-n}\, \delta^-_{i"} \delta^+_{j"} B_{\kappa_{i"} \kappa_{j"}}() )\right)=0
 \end{array} \end{equation}for $(\rho)_n=0$ and using the variable substitutions $q_1=p_1$ and $q_k-q_{k-1}=p_k$ for $k=2,\ldots n$. With both $M(p)$ and $B(p)$ multipliers of test functions, the concern is the generalized functions $\delta(q_n)$ and $\delta^{\pm}((q_i\!-\!q_{i-1})^2 \!-\!m_{\kappa_i}^2)$. The products of the energy-momentum conserving and mass shell delta functions are locally summable on the surface determined by the support of the delta functions, as shown in Appendix A. Application of the Riemann-Lebesque lemma for absolutely summable functions [\ref{titchmarsh}] establishes that (\ref{cluster-cond}) is satisfied as $r$ grows without bound. $\Upsilon(p)$ and the $U_k(p)$ are multipliers and preserve this argument.

The two-point function is connected and has light-cone singularities that are evaded by limiting the cluster decompositions to space-like translations.
 

\section*{Appendices}
\subsection*{A. The generalized function $\delta(p_1+p_2+\ldots p_n)\, (\delta^-)_k\, (\delta^+)_{k+1,n}$}

$\delta(p_1+p_2+\ldots p_n)\, (\delta^-)_k\, (\delta^+)_{k+1,n}$ are defined as generalized functions of $(p)_n \in {\bf R}^{nd}$ by summations over the surface with energy and momentum conserved, and energies on mass shells. The singularities of the measure on the surface $({\bf p})_n \in {\bf R}^{n(d-1)}$ induced by\begin{equation} \delta(P_k):= \delta(\omega_1 \ldots +\omega_k -\omega_{k+1} \ldots -\omega_n)\;\delta({\bf p}_1\!+\!{\bf p}_2\ldots \!+\!{\bf p}_n)\label{delta}\end{equation}are locally summable for $d\geq 3$.

With $({\bf p})_n$ constrained to the surface\begin{equation}\label{surface} {\bf p}_n=-{\bf p}_1\ldots -{\bf p}_{n-1},\end{equation}summation over the surface that satisfies $P_{k(0)}=0$ defines a generalized function except possibly for points on the surface with a vanishing gradient [\ref{gel1}].

Define $s_j$ by\[P_{k(0)}= \sum_{j=1}^n s_j \omega_j\]or $s_i:=1$ for $i\leq k$, and $s_i:=-1$ for $i>k$. The cases $k=0,n$ have no solutions to $P_{k(0)}=0$ and are not considered below. Then $s_1=-s_n=1$. On the indicated surface, the components of the gradient are\begin{equation} \renewcommand{\arraystretch}{2} \frac{\partial P_{k(0)}}{\partial p_{j(\ell)}} = s_j \frac{\ds p_{j(\ell)}}{\ds \omega_j}-s_n\frac{\ds p_{n(\ell)}}{\ds \omega_n} \label{grad}\end{equation}for $j = 1,\ldots n\!-\!1$ and $\ell = 1,\ldots d\!-\!1$. The components of ${\bf p}_j$ are designated $p_{j(1)},p_{j(2)},\ldots p_{j(d-1)}$. Summing squares provides that when the gradient vanishes\begin{equation}\label{gradvansh} \frac{\omega_j}{m_{\kappa_j}}= \frac{\omega_n}{m_{\kappa_n}}.\end{equation}
 
A neighborhood $V$ of those points in the surface (\ref{surface}) with a vanishing gradient is\[\frac{{\bf p}_j}{m_{\kappa_j}} =s_j \frac{{\bf p}_1}{m_{\kappa_1}} +{\bf e}_j\]for $j= 2,\ldots n\!-\!1$ with $\| {\bf e}_j \| < \epsilon$ arbitrarily small and $m_{\kappa_n}{\bf p}_1\approx - m_{\kappa_1} {\bf p}_n$. In $V$,\[\renewcommand{\arraystretch}{2} \begin{array}{rl} {\bf p}_n &=-{\ds \sum_{j=1}^{n-1}}{\bf p}_j\\
 &=-{\ds \sum_{j=1}^{n-1}}s_j{\ds \frac{m_{\kappa_j}}{m_{\kappa_1}}} {\bf p}_1-{\ds \sum_{j=2}^{n-1}} m_{\kappa_j} {\bf e}_j\\
 &= -{\ds \frac{m_{\kappa_n}}{m_{\kappa_1}}} {\bf p}_1 -{\ds \sum_{j=2}^{n-1}} m_{\kappa_j} {\bf e}_j -{\ds \sum_{j=1}^n} s_jm_{\kappa_j}{\ds \frac{ {\bf p}_1}{m_{\kappa_1}}}.\end{array}\]If\[\sum_{j=1}^n s_j m_{\kappa_j} \neq 0,\]then the unique solution to a vanishing gradient (\ref{grad}) has ${\bf p}_1=0$, $({\bf e})_{2,n-1}=0$. But in this case, $P_{k(0)} \neq 0$ when the gradient vanishes. The case of interest has\begin{equation}\label{mass-eq}\sum_{j=1}^n s_j m_{\kappa_j} = 0.\end{equation}In this case, the gradient vanishes when $({\bf e})_{2,n-1}=0$ and when the gradient vanishes, $P_{k(0)} = 0$ from (\ref{gradvansh}). In $V$,\[\omega_j \approx \frac{m_{\kappa_j}}{m_{\kappa_1}}\; \omega_1 + s_j m_{\kappa_j} \frac{{\bf p}_1\cdot {\bf e}_j}{\omega_1} +m_{\kappa_1}m_{\kappa_j}\frac{{\bf e}_j \cdot {\bf e}_j}{2\omega_1}-m_{\kappa_1}m_{\kappa_j}\frac{({\bf p}_1 \cdot {\bf e}_j)^2}{2\omega_1^3}\]to second order in small quantities and with\begin{equation}\label{error-eq}m_{\kappa_n}{\bf e}_n = -\sum_{\ell=2}^{n-1} m_{\ell}{\bf e}_{\ell}.\end{equation}

Within $V$, due to (\ref{mass-eq}) and (\ref{error-eq}),\[ \renewcommand{\arraystretch}{2} \begin{array}{rl} P_{k(0)} &\approx \frac{\ds m_{\kappa_1}}{\ds 2\omega_1^3}\; {\ds \sum_{j=2}^n} s_j m_{\kappa_j} \left( \omega_1^2\; {\bf e}_j \cdot {\bf e}_j -({\bf p}_1 \cdot {\bf e}_j)^2\right)\\
 &= \frac{\ds R^2}{\ds 2\omega_1^3}\;(a\omega_1^2-br_1^2) \end{array}\]with $r_1$ the length of ${\bf p}_1$,\[ \renewcommand{\arraystretch}{2} \begin{array}{rl} a R^2&:= m_{\kappa_1}{\ds \sum_{j=2}^n} s_j m_{\kappa_j} \left( {\bf e}_j \cdot {\bf e}_j\right)\\
b R^2&:= m_{\kappa_1}{\ds \sum_{j=2}^n} s_j m_{\kappa_j} ({\bf u}_1 \cdot {\bf e}_j)^2,\end{array}\]the unit vector ${\bf u}_1 :={\bf p}_1/ r_1$, and\[R^2:=\sum_{j=2}^{n-1} {\bf e}_j \cdot {\bf e}_j.\]This choice for polar coordinates has $R$ as the Euclidean length of $({\bf e})_{2,n-1}$ and $a,b$ are independent of $R$ consistent with the approximations in $V$.
 
Within $V$, the generalized function factors.\[ \renewcommand{\arraystretch}{2} \begin{array}{rl} \delta(P_{k(0)}) &= \delta(\frac{\ds R^2}{\ds 2\omega_1^3}\;(a\omega_1^2-br_1^2) )\\
 &= \frac{\ds 2\omega_1^3}{\ds R^2}\;\delta(a\omega_1^2-br_1^2) + \frac{\ds 2\omega_1^3}{\ds a\omega_1^2-br_1^2}\; \delta(R^2 ).\end{array}\]Since the generalized function is defined for $R>0$ and $a\omega_1^2-br_1^2$ is independent of $R$, $\delta(a\omega_1^2-br_1^2)$ is defined for $R=0$. Then, (\ref{delta}) defines a generalized function (without regularization) if\[\frac{\ds \omega_1^3}{\ds R^2}\]is locally summable. This is singular, diverging as $R^{-2}$ but $d\geq 3$ suffices since the Jacobian for the polar coordinates for $({\bf e})_{2,n-1}$ contributes $R^{(d-1)(n-2)-1}$ and $n\geq4$. The second term vanishes for $d\geq 3$.

\subsection*{B. Example realizations}
The constructions include scalar $M(p)=S(A)=D=1$.
 
A two component scalar field has $2\times 2$ $S(A)=1$, and\[ \renewcommand{\arraystretch}{1} \begin{array}{rl} 
 D &= \left( \begin{array}{cc} 0 & 1\\ 1 & 0\end{array} \right) \\
 M(p) &= \left( \begin{array}{cc} 0 & 1\\ 1 & 0\end{array} \right)=M(-p)^T. \end{array} \]This field has a symmetry associated with a conserved charge, $S_{\phi}M(p)S_{\phi}^T=M(p)$.\[ S_{\phi} = \left( \begin{array}{cc} e^{i\phi} & 0\\ 0 & e^{-i\phi}\end{array} \right)\]and $DS_{\phi}=\overline{S}_{\phi}D$.

An example with spin-1 bosons results from a real representation of the Lorentz group and a $d\times d$ $D=1$.\[ \renewcommand{\arraystretch}{1.25} \begin{array}{rl} 
 D &= 1\\
 S(A) &= \Lambda(A)\\
 M(p)_{jk} &= \frac{\ds p_{(j)} p_{(k)}}{\ds m^2} - g_{jk}\\
 &= C^*(p)C(p). \end{array} \]$g$ is the Minkowski metric. $M(p)=M(-p)^T$. In four dimensions with $m=1$ and $p=(E,p_x,p_y,p_z)$,\begin{equation} C^*(p) = \left( 
\begin{array}{cccc} \sqrt{E^2-1} & 0 & 0 & 0\\
 \frac{\ds p_xE}{\ds \sqrt{E^2-1}} & \sqrt{\frac{\ds p_y^2+p_z^2}{\ds E^2-1}} & 0 & 0\\
 \frac{\ds p_yE}{\ds \sqrt{E^2-1}} & \frac{\ds -p_xp_y}{\ds \sqrt{(p_y^2+p_z^2)(E^2-1)}} & \frac{\ds p_z}{\ds \sqrt{p_y^2+p_z^2}} & 0\\
 \frac{\ds p_zE}{\ds \sqrt{E^2-1}} & \frac{\ds -p_xp_z}{\ds \sqrt{(p_y^2+p_z^2)(E^2-1)}} & \frac{\ds -p_y}{\ds \sqrt{p_y^2+p_z^2}} & 0\end{array} \right).\end{equation}

An example with spin-1/2 fermions in $4$ dimensions uses a $4 \times 4$ reducible representation of the Lorentz group.\[ \renewcommand{\arraystretch}{1.25} \begin{array}{rl} 
 D &= \left( \begin{array}{cc} 0 & 1\\ 1 & 0\end{array} \right) \\
 S(A) &= \left( \begin{array}{cc} A & 0 \\ 0 & \overline{A}\end{array} \right)\\
 M(p) &= \left( \begin{array}{cc} 0 & P\\ P^T & 0\end{array} \right)=-M(-p)^T \\
 P &= P^* = \left( \begin{array}{cc} E+p_z & p_x-ip_y \\ p_x+ip_y & E-p_z\end{array} \right)\end{array} \]and $A \in \mbox{SL}(2)$.\begin{equation} 
D M(p) = C^*(p) C(p) = \left( \begin{array}{cc} c^T & 0\\ 0& c^*\end{array} \right)\left( \begin{array}{cc} \overline{c}& 0\\ 0& c\end{array} \right)\end{equation}is a nonnegative matrix for $p$ within the forward cone and\begin{equation} c(p) = \left( \begin{array}{cc} \sqrt{E+p_z} & \frac{\ds p_x+ip_y}{\ds \sqrt{E+p_z}}\\ 0& \frac{\ds m}{\ds \sqrt{E+p_z}}\end{array} \right). \end{equation}This field also has a symmetry $S_{\phi}M(p)S_{\phi}^T=M(p)$ for\[ S_{\phi} = \left( \begin{array}{cc} \phi & 0\\ 0 & \overline{\phi}\end{array} \right)\qquad \phi = \left( \begin{array}{cc} e^{i\phi} & 0\\ 0 & e^{i\phi}\end{array} \right)\]and $DS_{\phi}=\overline{S}_{\phi}D$.

In these examples, the elements of $M(p)$ are multinomials in the energy-momentum components and this persists for greater spins [\ref{bogo},\ref{gel5}]. Additional $M(p),S(A),D$ result from compositions in Kronecker (direct, or tensor) products and direct sums. Given a real orthogonal transform $O$ and a construction $M(p),S(A),D$, then $M(p)'=OM(p)O^T$, $S(A)'= OS(A)O^T$, $D'= ODO^T$ also satisfy (\ref{matcond}), (\ref{matcond0}), and (\ref{matcond2}). Composing models, distinct $U_n(p),\Upsilon(p)$ can apply to each particle species.

\section*{References}
\begin{enumerate}
\item \label{wight} A.S.~Wightman, ``Quantum Field Theory in Terms of Vacuum Expectation Values'', {\em Phys.~Rev.}, vol.~101, 1956, p.~860.
\item \label{pct} R.F.~Streater and A.S.~Wightman, {\em PCT, Spin and Statistics, and All That}, Reading, MA: W.A.~Benjamin, 1964.
\item \label{bogo} N.N.~Bogolubov, A.A.~Logunov, and I.T.~Todorov, {\em Introduction to Axiomatic Quantum Field Theory}, trans.~by Stephen Fulling and Ludmilla Popova, Reading, MA: W.A.~Benjamin, 1975.
\item \label{borchers} H.J.~Borchers, ``On the structure of the algebra of field operators'', {\em Nuovo Cimento}, Vol.~24, 1962, p.~214.
\item \label{locality} H.~Baumg\"{a}rtel, and M.~Wollenberg, ``A Class of Nontrivial Weakly Local Massive Wightman Fields with Interpolating Properties'', {\em Commun. Math. Phys.}, Vol.~94, 1984, p.~331.
\item \label{lechner} D.~Buchholz, L.~Lechner and S.J.~Summers, ``Warped Convolutions, Rieffel Deformations and the Construction of Quantum Field Theories'', arXiv:1005.2656v1 [math-ph], May 15, 2010.
\item \label{gel1} I.M.~Gelfand, and G.E.~Shilov, {\em Generalized Functions, Vol.~1}, trans.~by E.~Saletan, New York, NY: Academic Press, 1964.
\item \label{steinmann} O. Steinmann, ``Structure of the Two-Point Function'', {\em Journal of Math. Phys.}, Vol. 4, 1963, p. 583.
\item \label{gel5} I.M.~Gel'fand, M.I.~Graev, and N.Ya.~Vilenkin, {\em Generalized Functions}, Vol.~5, trans.~E.~Saletan, New York, NY: Academic Press, 1966.
\item \label{bjdrell2} J.D.~Bjorken and S.D.~Drell, {\em Relativistic Quantum Fields}, New York, NY: McGraw-Hill, 1965.
\item \label{wigner} T.D.~Newton and E.P.~Wigner, ``Localized States for Elementary Systems'', {\em Rev. Modern Phys.}, Vol.~21, 1949, p.~400.
\item \label{segal} I.E.~Segal and R.W,~Goodman, ``Anti-locality of certain Lorentz-invariant operators'', {\em Journal of Mathematics and Mechanics}, Vol.~14, 1965, p.~629.
\item \label{mund} J.~Mund, ``An Algebraic Jost-Schroer Theorem for Massive Theories'', arXiv: 1012.1454v3 [hep-ph], Dec. 2010.
\item \label{type3} J.~Yngvason, ``The Role of Type III Factors on Quantum Field Theory'', arXiv:math-ph/0411058v2 [math-ph], Dec. 2004.
\item \label{yngvason} J. Yngvason, ``The Borchers-Uhlmann Algebra and its Descendants'', G\"{o}ttingen, July 29, 2009, www.lqp.uni-goettingen.de/events/aqft50/slides/1-4-Yngvason.pdf.
\item \label{summers} G.~Lechner, ``Deformations of quantum field theories and integrable models'', arXiv:\-1104.\-1948v2 [math-ph], May 1, 2011.
\item \label{ewg} B.S.~DeWitt, H.~Everett III, N.~Graham, J.A.~Wheeler republished in {\em The Many-worlds Interpretation of Quantum Mechanics}, ed.~B.S.~DeWitt, N. Graham, Princeton, NJ: Princeton University Press, 1973.
\item \label{entangle} R.~Horodecki, P.~Horodecki, M.~Horodecki, K.~Horodecki. ``Quantum entanglement'', {\em Rev. Mod. Phys.}, Vol.~81 (2), p. 865, arXiv:quant-ph/0702225, 2007.
\item \label{gel4} I.M.~Gel'fand, and N.Ya.~Vilenkin, {\em Generalized Functions}, Vol.~4, trans.~A.~Feinstein, New York, NY: Academic Press, 1964.
\item \label{schrodinger} E. Schr\"{o}dinger, ``Der stetige Übergang von der Mikro-zur Makromechanik'', {\em Die Naturwissenschaften}, Vol.~14. Issue~28, 1926, p.~664.
\item \label{agw} S.~Albeverio, H.~Gottschalk, and J.-L.~Wu, ``Convoluted generalized white noise, Schwinger functions and their analytic continuation to Wightman functions'', {\em Rev. Math. Phys.}, Vol.~8, 1996, pg.~763-817.
\item \label{iqf} G.E.~Johnson, ``Interacting quantum fields'', {\em Rev.~Math.~Phys.}, Vol.~11, 1999, p.~881 and errata, {\em Rev.~Math.~Phys.}, Vol.~12, 2000, p.~687. (The constructed Euclidean region semi-norm is insufficient to imply a physical region semi-norm for a $*$-involutive algebra and consequent Hilbert space field operators. The fields are operations in the topological space of function sequences.)
\item \label{gel2} I.M.~Gel'fand, and G.E.~Shilov, {\em Generalized Functions}, Vol.~2, trans.~M.D.~Friedman, A.~Feinstein, and C.P.~Peltzer, New York, NY: Academic Press, 1968.
\item \label{emch} G.E.~Emch, {\em Algebraic Methods in Statistical Mechanics and Quantum Field Theory}, New York, NY: Wiley-Interscience, 1972.
\item \label{cook} J.M.~Cook, ``The Mathematics of Second Quantization'', {\em Trans.~Am.~Math.~Soc.}, Vol.~74, 1953, pg.~222.
\item \label{horn} R.A. Horn, and C.R. Johnson, {\em Matrix Analysis}, Cambridge: Cambridge University Press, 1985.
\item \label{combin} M. Hardy, ``Combinatorics of Partial Derivatives'', {\em The Electronic Journal of Combinatorics}, Vol. 13, \#R1, 2006.
\item \label{heger} G.C. Hegerfeldt, ``Prime Field Decompositions and Infinitely Divisible States on Borchers' Tensor Algebra'', {\em Commun. Math. Phys.}, Vol.~45, 1975, p.~137.
\item \label{vonN} J.~von Neumann, {\em Mathematical Foundations of Quantum Mechanics}, Princeton, NJ: Princeton University Press, 1955.
\item \label{ruelle1} D. Ruelle, {\em Statistical Mechanics, Rigorous Results}, Reading, MA: W.A.~Benjamin, 1969.
\item \label{titchmarsh} E.C.~Titchmarsh, {\em Introduction to the Theory of Fourier Integrals}, Oxford, The Oxford University Press, 1948.
\end{enumerate}
\end{document}